\documentclass[final]{IEEEtran}

\usepackage{subcaption}
\usepackage{graphicx}  

\usepackage{cuted} 
\usepackage{amsmath}
\usepackage{amssymb}
\usepackage{url}
\usepackage{hyperref}%
\hypersetup{colorlinks=true, linkcolor=blue, breaklinks=true, urlcolor=blue, citecolor=blue}
\usepackage{enumitem}
\usepackage{mleftright}

\usepackage{tikz}
\usetikzlibrary{matrix,positioning,calc}

\usepackage{todonotes}
\usepackage{setspace}
\usepackage{comment}
\usepackage{amsthm}

\newcommand{\tr}{\operatorname{tr}}

\newcommand{\diag}{\operatorname{diag}}
\newcommand{\st}{\, | \,}
\newcommand{\be}{\begin{equation}}
\newcommand{\ee}{\end{equation}}
\newcommand{\R}{\mathbb R}
\newcommand{\C}{\mathbb C}

\newcommand{\blkotimes}{\overline{\otimes}}

\newcommand{\until}[1]{\{1,\dots, #1\}}
\newcommand{\mcG}{\mathcal{G}}
\newcommand{\mcC}{\mathcal{C}}
\usepackage{algorithm2e,algorithmic}
\newcommand{\gain}[1]{\operatorname{gain}(#1)}
\renewcommand{\gain}[1]{\gamma_{#1}}
\newcommand{\ds}{\displaystyle}

\usepackage{tikz}
\usepackage{tikz-cd}
\usetikzlibrary{automata, positioning}
\usetikzlibrary{shapes,backgrounds}
\usetikzlibrary{positioning, shapes.geometric}

\usetikzlibrary{decorations.pathmorphing, arrows.meta}

\DeclareSymbolFont{bbold}{U}{bbold}{m}{n}
\DeclareSymbolFontAlphabet{\mathbbold}{bbold}

\newtheorem{Theorem}{Theorem}
\newtheorem{Lemma}[Theorem]{Lemma}
\newtheorem{Proposition}[Theorem]{Proposition}
\newtheorem{Corollary}[Theorem]{Corollary}
\newtheorem{Definition}[Theorem]{Definition}
\newtheorem{Remark}[Theorem]{Remark}

\newtheorem{example}[Theorem]{Example}
\AtBeginEnvironment{example}{%
  \pushQED{\qed}%
}
\AtEndEnvironment{example}{\popQED\endexample}

\title{A sufficient condition for 2-contraction \\ of  a feedback interconnection}

\author{Ron Ofir and Francesco Bullo and  Michael Margaliot\thanks{RO and MM (michaelm@tauex.tau.ac.il) are with the School of Elec. Eng., Tel Aviv University, 69978, Israel.
The research of RO is partly supported by a  Khazanov Scholarship. The research of MM is partly supported by a  research grant from the Israeli Science Foundation~(ISF). FB (bullo@ucsb.edu) is with the Center for Control, Dynamical Systems, and Computation, UC Santa Barbara, Santa Barbara, CA 93106 USA. FB is partly supported by AFOSR grant FA9550-22-1-0059. An abridged version of this paper will be presented at the 60th Allerton Conference, to be held in Sept. 25-27,  2024, at Urbana, Illinois.}}

\begin{document}
\maketitle

\begin{abstract}
Multistationarity -- the existence of multiple equilibrium points -- is a common phenomenon in  dynamical systems from a variety of fields, including neuroscience, opinion dynamics, systems biology, and power systems. A recently proposed generalization of contraction theory, called $k$-contraction, is a promising approach for analyzing the asymptotic behaviour of multistationary systems. In particular, all bounded trajectories of a time-invariant 2-contracting system converge to an equilibrium point, but the system may have multiple equilibrium points where more than one is locally stable. An important challenge is to study $k$-contraction in
large-scale interconnected  systems. Inspired by a recent small-gain theorem for 2-contraction by Angeli et al., we derive a new sufficient condition for 2-contraction of  a feedback interconnection of two nonlinear dynamical systems.
Our condition is based on (i) deriving new formulas for the 2-multiplicative [2-additive] compound of block matrices using block Kronecker products [sums], (ii) a hierarchical approach for proving standard contraction, and (iii) a network small-gain theorem for Metzler matrices.
We demonstrate our results by deriving  a simple sufficient condition for $2$-contraction in a network of FitzHugh-Nagumo neurons.
\end{abstract}


\section{Introduction}

Many dynamical models exhibit multistationarity, i.e. the coexistence of multiple equilibrium points.  Examples include models from  systems biology~\cite{detect_multi_stab}, nonlinear opinion dynamics~\cite{Bizyaeva2024NonlinOpinion}, dynamic   neural networks~\cite{Cheng2006MultistableRNNs}, mathematical epidemiology~\cite{Kiss2017Epidemics}, 
and power systems~\cite{Nguyen2015VoltMultStab,Delabays2022MultistabGrids}. A promising 
approach for the asymptotic analysis of multistationary systems is a recently proposed generalization of contraction theory, called $k$-contraction~\cite{Wu2022kcontraction}, that is based on the seminal work 
of Muldowney and his colleagues~\cite{Muldowney1990CompoundODE,Li1995SmithAutoConv}.

Roughly speaking, a system is called $k$-contracting if its flow contracts the volume of $k$-dimensional bodies at an exponential rate~\cite{Wu2022kcontraction}. Since line segments are 1-dimensional bodies,   
1-contracting systems are conventional contracting systems~\cite{Lohmiller1998Contraction,Aminzare2014Contraction}. Unlike standard contracting systems, which are globally asymptotically stable~\cite{Bullo2024ContractionBook}, and so cannot be multistationary, $k$-contracting systems may have multiple equilibrium points, while still exhibiting a well-ordered asymptotic behavior. In particular, any bounded trajectory of a time-invariant 2-contracting system  converges to the set of equilibria.

The theory of $k$-contraction uses tools from multilinear algebra known as multiplicative and additive compound matrices. Compound matrices have found applications
in several mathematical disciplines, including graph theory~\cite{Fiedler1998AdditiveCompoundGraphs,Bapat2013CompoundTrees}, linear algebra~\cite{London1976DerivationsDiffEq}, and differential equations~\cite{Schwarz1970TPDS}. Compound  matrices also have many applications in systems and control theory~\cite{BarShalom2023CompoundTutorial}. These include $k$-contraction~\cite{Wu2022kcontraction} and the related concept of~$\alpha$-contraction~\cite{Wu2022AlphaContraction}, with~$\alpha$ a real number, as well as various 
generalizations of positive and cooperative  systems~\cite{Weiss2021kpositive,
Weiss2021IsMySystemkcoop, Grussler2022HankelkPositive,Grussler2022VariationDiminishing,rami_3D}. These different generalizations are based on different properties of compound matrices, in particular their geometric interpretation for $k$- and $\alpha$-contraction, and their 
effect on the number of 
   sign variations in a vector for the generalizations of cooperative systems.
   In all these generalizations of  continuous-time dynamical systems a basic tool is 
   the additive compound of the Jacobian of the vector field describing the  dynamical system.

Studying the additive compound directly may  be challenging for at least two reasons.
First, 
the dimensions of the $k$-additive compound of an~$n\times n$ Jacobian 
matrix 
are~$\binom{n}{k}\times \binom{n}{k}$. For example,  the~$2$-additive compound of a~$6$-dimensional dynamical system is a matrix with dimensions~$\binom{6}{2}\times\binom{6}{2}=15\times 15$. Second, 
any structure in the Jacobian itself might not be reflected in an obvious way  in the $k$-compounds of the Jacobian. For this reason, several recent works have considered alternative methods for proving $k$-contraction. These include a sufficient condition for $k$-contraction based only on the Jacobian and its trace~\cite{DalinOBSOBM2022Duality}, and using  LMI-based sufficient conditions~\cite{Cecilia2023GenLyapkcont,Zoboli2023LMIkcont}. For more recent work on $k$-contraction, see also~\cite{k-singular,partial-and-k-contraction,ANGELI2024152}.

Consider a system composed of several interconnected sub-systems. The Jacobian of this system will have a  block structure, with the diagonal blocks corresponding to the sub-systems, and the off-diagonal blocks representing the interconnections between the sub-systems.   A natural ides is to analyze~$k$-contraction in such a system  based on analyzing the $k$-compounds of the Jacobians of the sub-systems, thus simplifying the analysis of the overall system. Recent results in this direction include sufficient conditions for $k$-contraction in:
    (1)~the series interconnection  of several  sub-systems~\cite{OfirMLS2021SeriesCDC,OfirMLS2021SeriesTAC};
    (2)~the feedback interconnection of an  LTI sub-system and a memoryless nonlinearity  (i.e., a Lurie system)~\cite{OfirOM2023LurieIFAC,OfirOM2024LurieAuto};
    and (3)~a small-gain theorem for 2-contraction in the feedback interconnection of two sub-systems~\cite{Angeli2023smallgain}.

In particular, the small-gain theorem derived in~\cite{Angeli2023smallgain} is based on an explicit expression for the 2-additive compound of a block matrix. However, this explicit expression is non-trivial, and the resulting sufficient condition for 2-contraction of the interconnected feedback   system is difficult to interpret in terms of the sub-systems themselves.  

Our main results include the following.  First, we  derive  a novel  expression for the $2$-multiplicative compound of a block matrix. We believe that this expression will be useful in the analysis of $2$-contractive and~$2$-cooperative dynamical systems. We demonstrate this by showing that this expression immediately provides an expression 
for the $2$-additive compound of a block matrix that is simpler than the expression  recently obtained  in~\cite{Angeli2023smallgain} using  a different approach. 
Second, we combine  the new expression 
for the $2$-additive compound of a block matrix and a hierarchic approach to proving standard contraction~\cite{Russo2013NetworkContractive} 
to derive a novel  sufficient condition for~$2$-contraction in non-linear feedback systems.
We then adopt the network small-gain theorem for Metzler matrices, recently developed in~\cite{Xiaoming2021GraphMetzler}, to render the final sufficient condition even more explicit.
Our conditions are in line with the theory on large-scale systems, as are based in a clear way on the $k$-contraction properties of the sub-systems and the norms of the interconnections. 
We demonstrate the theoretical results by deriving a simple and easy to verify sufficient condition for $2$-contraction in a network of FitzHugh-Nagumo neurons. 
Our results apply to  an arbitrary interconnection matrix, and  simplify further  for  the typical case of  a diffusively-coupled network, that is, when the connection matrix is a Laplacian.

We use standard notation. For a matrix~$A \in \R^{n \times m}$,~$a_{ij}$ denotes the~$(i,j)$ entry of~$A$, and~$A^T$ denotes the   transpose of~$A$.    
A square matrix~$A$ is called Metzler if~$a_{ij}\geq0 $ for  all~$i\not =j$. The determinant  [trace] of~$A $ is  denoted by~$\det(A)$ [$\tr(A)$]. The $n\times n$ identity matrix is denoted by~$I_n$. 

  We use~$|\cdot|:\R^n\to\R_{\geq 0}$ to denote  a vector norm, and~$\|\cdot\|:\R^{n\times n}\to \R_{\geq 0}$ to denote the
  induced matrix norm. 
Recall that a norm~$|\cdot|:\R^n\to\R_{\geq 0}$ is called monotonic if~$x_i\geq y_i$ for all~$ i \in \{1,\dots,n\}$ 
implies that~$|x|\geq|y|$~\cite{Bauer1961MonoNorms}. 

The remainder of this paper
is organized as follows. The next section reviews known definitions and results that are used later on. Section~\ref{sec:main}  states the main results, the proof of which is placed
in Section~\ref{sec:proof}. Section~\ref{sec:app} presents an application to a dynamical
neural network, and the last section concludes and describes possible directions for further research.

\section{Preliminaries}\label{sec:prelim}

\subsection{Compound matrices}
The theory of $k$-contraction is based on an important notion from multilinear algebra called compound matrices. In this section, we review the multiplicative and additive compounds of a (real) matrix.   More details   and applications to systems and control theory can be found in the recent tutorial paper~\cite{BarShalom2023CompoundTutorial}.

Recall that a $k$ minor of a matrix~$A$ (or minor of order $k \times k$) is the determinant of some~$k\times k$ submatrix of~$A$. 

\begin{Definition}\label{eq:def_multi}
Let~$A\in\R^{n\times m}$. Fix~$k\in\{1,\dots,\min\{n,m\}\}$.
    The \emph{$k$-multiplicative compound} of~$A$, denoted~$A^{(k)}$, is the~$\binom{n}{k}\times\binom{m}{k}$ matrix that contains all the~$k$ minors of~$A$ in lexicographic order.
\end{Definition}

\begin{example}
    The~$k$-multiplicative compound of 
    the matrix~$I_n$ is~$I_r$, where~$r:=\binom{n}{k}$. 
\end{example}

\begin{example}
    Suppose that~$n=m=3$ and~$k=2$. Then 
    the 2-multiplicative compound of~$A$ is the~$3\times 3$ matrix:
\begin{align*}
A^{(2)} &= 
    \begin{bmatrix}
        \det\mleft( \begin{smallmatrix} a_{11} & a_{12} \\ a_{21} & a_{22} \end{smallmatrix} \mright)  &
        \det\mleft( \begin{smallmatrix} a_{11} & a_{13} \\ a_{21} & a_{23} \end{smallmatrix} \mright)  &
        \det\mleft( \begin{smallmatrix} a_{12} & a_{13} \\ a_{22} & a_{23} \end{smallmatrix} \mright)  \\
        \det\mleft( \begin{smallmatrix} a_{11} & a_{12} \\ a_{31} & a_{32} \end{smallmatrix} \mright)  &
        \det\mleft( \begin{smallmatrix} a_{11} & a_{13} \\ a_{31} & a_{33} \end{smallmatrix} \mright)  &
        \det\mleft( \begin{smallmatrix} a_{12} & a_{13} \\ a_{32} & a_{33} \end{smallmatrix} \mright)  \\
        \det\mleft( \begin{smallmatrix} a_{21} & a_{22} \\ a_{31} & a_{32} \end{smallmatrix} \mright)  &
        \det\mleft( \begin{smallmatrix} a_{21} & a_{23} \\ a_{31} & a_{33} \end{smallmatrix} \mright)  &
        \det\mleft( \begin{smallmatrix} a_{22} & a_{23} \\ a_{32} & a_{33} \end{smallmatrix} \mright)
    \end{bmatrix}.
\end{align*}
\end{example}

\begin{Definition}\label{def:add}
    The \emph{$k$-additive compound} of~$A \in \R^{n \times n}$, denoted~$A^{[k]}$, is the~$\binom{n}{k}\times\binom{n}{k}$ matrix defined by
    \begin{equation}\label{eq:def_add}
        A^{[k]} := \frac{d}{d\varepsilon} \left( I_n + \varepsilon A \right)^{(k)}|_{\varepsilon=0}.
    \end{equation}
\end{Definition}

In other words, the Taylor series of~$(I_n+\varepsilon A)^{(k)}$ is
\begin{align}\label{eq:taylor_series}
(I_n+\varepsilon A)^{(k)}  &=  (I_n)^{(k)} +\varepsilon A^{[k]} +o(\varepsilon)\nonumber\\
&= I_r + \varepsilon A^{[k]} +o(\varepsilon),
\end{align}
with~$r:=\binom{n}{k}$. 

\begin{example}
Consider  the $k$-additive compound of the identity  matrix~$I_n$. By~\eqref{eq:def_add},
\begin{align*}
    I_n^{[k]} & = \frac{d}{d\varepsilon} (I_n+\varepsilon I_n )^{(k)}|_{\varepsilon=0}\\
    &= \frac{d}{d\varepsilon} (1+\varepsilon )^k I_n^{(k)}|_{\varepsilon=0}\\
    &=k I_r,
\end{align*}
where~$r:=\binom{n}{k}$.
\end{example}
\begin{example}
To explain the geometric importance of these compounds, suppose  that~$A$ is~$3\times 3$, and
diagonal, that is,~$A=\diag(a_{11},a_{22},a_{33})$. Then 
\[
A^{(2)}  = 
    \begin{bmatrix} a_{11}a_{22} &0&0\\
    0& a_{11}a_{33}&0\\
    0&0&a_{22}a_{33}  
\end{bmatrix}, 
\]
and
\[
A^{[2]}  = 
    \begin{bmatrix} a_{11}+a_{22} &0&0\\
    0& a_{11}+a_{33}&0\\
    0&0&a_{22}+a_{33}  
\end{bmatrix}.
\]
Suppose  that~$x:[0,\infty)\to\R^3$ evolves according to
\be \label{eq:lti3}
\dot x (t)=Ax(t)=\diag(a_{11},a_{22},a_{33}) x(t) .
\ee
Let~$S_0:=[0,1]\times[0,1]\times 0$ and let~$S(t):=\{x(t)\st x(0)\in S_0\} $, that is, we consider the evolution of the 2D   unit square~$S_0$ under the LTI~\eqref{eq:lti3}. 
Then~$S(t)=[0,\exp(a_{11}t )]\times [0,\exp(a_{22}t)]\times 0$, so the area of~$S(t)$ is~$\exp((a_{11}+a_{22})t)$. 
More generally, it
is straightforward to verify
that any two-dimensional body shrinks  under the evolution of the LTI~\eqref{eq:lti3} 
iff~$A^{[2]}$ is Hurwitz.  
\end{example}

Definition~\ref{eq:def_multi} implies that~$\left(A^{(k)}\right)^T = (A^T)^{(k)}$, $A^{(1)} = A$, and if~$n=m$ then~$A^{(n)} = \det(A)$. The name multiplicative compound follows from the Cauchy-Binet Theorem, which states that \[
(AB)^{(k)} = A^{(k)} B^{(k)}
\]
for any $A \in \R^{n \times p}, B \in \R^{p \times m}$ and~$k \in \{1,\dots,\min\{n,p,m\}\}$. The Cauchy-Binet Theorem has several important implications. First, if~$A\in\R^{n\times n}$ is  non-singular 
then~$(A^{-1})^{(k)}=\left(A^{(k)}\right)^{-1}$. Also, if~$T,A\in\R^{n\times n}$ and~$T$ is non-singular then
\be\label{eq:TAT}
(TAT^{-1})^{(k)}=T ^{(k)} A ^{(k)} (T^{(k)})^{-1}. 
\ee
 
Definition~\ref{def:add}  implies that~$A^{[1]} = A$,  and~$A^{[n]} = \tr(A)$. For any~$\alpha \in \R$ and $A,B \in \R^{n \times n}$, we have that~$(\alpha A)^{[k]} = \alpha A^{[k]}$, and~$(A + B)^{[k]} = A^{[k]} + B^{[k]}$. Furthermore,   for any~$U \in \R^{m \times n}, V \in \R^{n \times m}$ and~$k \in \{1,\dots,\min\{n,m\}\}$, we have  
\begin{equation}
    (UAV)^{[k]} = U^{(k)} A^{[k]} V^{(k)}.
\end{equation}

One reason for the  importance of the~$k$-additive compound in dynamical systems follows from the fact that
\[
\frac{d}{dt} (\exp(At))^{(k)}=A^{[k]} (\exp(At))^{(k)}.
\]
In other words, the transition matrix~$(\exp(At))^{(k)}$, that describes how $k$-dimensional bodies evolve under the LTI~$\dot x=Ax$,   satisfies a linear differential equation with the matrix~$A^{[k]}$.
Hence,    the volume of any~$k$-dimensional body converges to zero at an exponential rate iff~$A^{[k]}$ is Hurwitz.

The next result (see, e.g.~\cite{Schwarz1970TPDS}) provides an explicit expression for the~$k$-additive compound. Recall that any entry of~$A^{(k)}$ is a minor of~$A$ corresponding to rows~$\alpha=\{i_1,\dots,i_k\}$ and columns~$\beta=\{j_1,\dots,j_k\}$,
with~$1\leq i_1<\dots<i_k\leq n$, 
and~$1\leq j_1<\dots<j_k\leq n$.
Thus, it is natural to index the entries of~$A^{(k)}$  and~$A^{[k]}$  using~$\alpha,\beta$.

\begin{Proposition}
    \label{prop:Explicit_exp_A_k} 
 Let~$A\in\R^{n\times n}$, and fix~$k\in\{1,\dots,n\}$. 
 The entry of~$A^{[k]}$ corresponding to~$(\alpha,\beta)$ is
equal to:
\begin{enumerate}
    \item $\sum_{\ell=1}^{k} a_{ i_{\ell} i_{\ell} }$, if  $i_{\ell} = j_{\ell}$  for all $\ell \in \{ 1,\hdots,k \}$;
    \item $(-1)^{\ell +m}
    a_{i_{\ell} j_{m}} $, 
      if all the indices  in $ \alpha  $  and $ \beta$   agree,  except  for   
   a single index $ i_{\ell} \ne j_m$; and
    \item $0$, otherwise.
\end{enumerate}
\end{Proposition}
In particular, this implies that any diagonal entry of~$A^{[k]}$ is the sum of~$k$ diagonal entries in~$A$,
and any off-diagonal entry is either zero or an entry of~$A$ multiplied  by either  plus or minus one. 

\subsection{$k$-contraction  }
Consider a nonlinear time-varying dynamical system
\begin{equation}\label{eq:nonlin}
    \dot x = f(t,x),
\end{equation}
with $f$   a $C^1$ function.
We assume throughout 
that there exists a convex state space~$\Omega \subseteq  \R^n$ such that for any~$x(0)\in\Omega$  the nonlinear system admits a unique solution~$x(t,x_0)\in\Omega$ for all~$t\geq 0 $. 
Let~$J(t,x) := \frac{\partial f}{\partial x}(t,x)$ denote the Jacobian of the vector field with respect to~$x$. 

Let~$|\cdot| : \R^n \to \R_{\geq 0}$ denote a vector norm, $\|A\|: = \max_{|x|=1} |Ax|$   the induced matrix norm, and $\mu(A) := \lim_{h \to 0^+} \frac{\|I + hA\| - 1}{h}$   the induced matrix measure  (also known as the logarithmic norm or log-norm). For example, the matrix measure corresponding to the~$L_2$ norm is~$\mu_2(A)=\frac{1}{2}\lambda_{\max}(A+A^T)$, where for a symmetric matrix~$S$, $\lambda_{\max}(S)$ is the maximal eigenvalue of~$S$.

We say that~\eqref{eq:nonlin} is (infinitesimally) $k$-contracting with respect to the norm~$|\cdot|$ and with rate~$\eta > 0$ if~$\mu(J^{[k]}(t,x)) \le -\eta$ for all $t \ge 0$ and~$ x \in \Omega$.
Here~$J^{[k]}(t,x) :=(J(t,x) )^{[k]}$
 is the $k$-additive compound of the Jacobian. 
Note that since~$J^{[1]}(t,x) = J(t,x)$,   this definition reduces to the standard condition for (infinitesimal) contraction in the case~$k=1$.

The main result for $k$-contracting systems, with $k>1$, is that any bounded 
solution of a time-invariant
2-contracting system converges to the set of equilibria (which may contain more than a single point). In this respect,  the system is ``globally multistable''.

\subsection{Kronecker products and sums}\label{subsec:kron}
The Jacobian of a feedback system is  a block matrix. Thus, analyzing~$k$-contraction  of a feedback systems requires computing the~$k$-compounds of a block matrix. We address this by  relating the multiplicative [additive] compound to a Kronecker product [sum], and then applying results on the Kronecker product of block matrices.

We briefly review the Kronecker product and sum,   see~\cite[Ch.~4]{HornJohnson1991Topics} for more details and proofs. 
 Fix~$A \in \R^{n \times m}$, $B \in \R^{r \times q}$. The Kronecker product of~$A$ and~$B$ is an~$(nr) \times (mq)$ matrix containing all products of every entry of~$A$ with every entry of~$B$:
\begin{equation*}
    A \otimes B := \begin{bmatrix}
        a_{11} B & \cdots & a_{1m} B \\
        \vdots & \ddots & \vdots \\
        a_{n1} B & \cdots & a_{nm} B
    \end{bmatrix}.
\end{equation*}
 Transposition is  distributive over the Kronecker product, that is,
 \[
 (A\otimes B)^T=A^T\otimes B^T. 
 \]
 The Kronecker
 product satisfies the following mixed-product property: if~$A,B,C,D$ are matrices  such that the products~$AC$ and~$BD$ are well-defined then
\begin{equation}\label{eq:mixed_prod}
    (A \otimes B)(C \otimes D) = (AC) \otimes (BD).
\end{equation}
The  ``$k$th Kronecker power''
is defined  
by
\begin{align*}
A^{\otimes k}:=  A \otimes A \otimes \dots   \otimes A,
\end{align*}
where the product includes~$k$ terms.

The Kronecker sum of two square matrices~$A \in \R^{n \times n}$ and~$B \in \R^{m \times m}$ is   
\begin{equation*}
    A \oplus B := A \otimes I_m + I_n  \otimes B. 
\end{equation*}
Note that~$A \oplus B \in \R^{(nm)\times(nm)}$.
For~$A\in\R^{n\times n}$, 
the  ``$k$th Kronecker sum'' of~$A$ is defined   by
\begin{align}\label{eq:def_k_kron_sum}
A^{\oplus k} &:=   A\otimes I_n \otimes \dots \otimes I_n+ 
   I_n\otimes A  \otimes I_n   \otimes\dots \otimes I_n\nonumber\\&+
\dots +
  I_n \otimes \dots \otimes I_n \otimes A, 
\end{align}
where every product includes~$k$ terms. 
For example,~$A^{\oplus 1}=A$, and~$A^{\oplus 2}= A\oplus A$.

Define the matrix~$H_{n,m}\in\R^{( nm)\times( nm)}$ by
\be\label{eq:def_H_permu}
H_{n,m}:=\sum_{i=1}^n\sum_{j=1}^m (E_{i,j}\otimes E_{i,j}^T),
\ee
where~$E_{i,j} \in\{0,1\}^{n\times m}$ is a matrix with 
   entry one at position~$(i,j)$ and all other entries are zero. Then~$H_{n,m}$ is a permutation matrix and for any 
$A\in\R^{n \times r} $ and~$B\in\R^{m \times q}$, 
we have
\be\label{eq:commute_h_mat}
A \otimes B=H_{n,m} (B\otimes A) H_{r,q}^T. 
\ee
In other words, the matrices~$H_{i,j}$ allow  to commute the order in a Kronecker product. 
For example, if~$A$ and~$B$ are vectors, that is,~$r=q=1$ then~\eqref{eq:commute_h_mat} becomes
\be\label{eq:a_otimes_b}
a \otimes b = H_{n,m} (b\otimes a) H_{1,1}^T = H_{n,m} (b\otimes a). 
\ee
Note that~\eqref{eq:def_H_permu} implies that
\begin{align}\label{eq:H
_mat_tanaspose}
H_{n,m}^T&=\sum_{i=1}^n\sum_{j=1}^m (E_{i,j}\otimes E_{i,j}^T)^T \nonumber\\
&=\sum_{i=1}^n\sum_{j=1}^m (E_{i,j}^T\otimes E_{i,j})\nonumber\\
&=\sum_{i=1}^n\sum_{j=1}^m (E_{j,i}\otimes E_{j,i}^T)\nonumber\\
&=H_{m,n}.
\end{align}
In particular,~$H_{n,n}$ is symmetric.  Combining  this with~\eqref{eq:commute_h_mat} implies that 
for any $A \in \R^{n \times n}$ and~$ B \in \R^{n \times m}$, we have 
\begin{align}
\label{eq:A_MUL_B_REP}
B\otimes A&=H_{n,n}(A\otimes B) H_{m,n}^T.
\end{align} 
 
It follows from the singular value decomposition and the mixed-product property that~$\|A \otimes B\|_2 = \|A\|_2 \|B\|_2$. Also, since~$A \oplus B + (A \oplus B)^T = (A + A^T) \oplus (B + B^T)$, we have that $\mu_2(A \oplus B) = \mu_2(A) + \mu_2(B)$. The next result shows that 
these equations   hold  for any~$L_p$ norm.
\begin{Lemma}[\cite{Lancaster1972normstensors},\cite{Wu2022AlphaContraction}]\label{lem:AotimesB_pnorm}
    Fix~$p \in [1,\infty]$,  $A \in \R^{n \times m}$, and~$ B \in \R^{r \times q}$. Then
    \begin{align*}
        \|A \otimes B\|_p &= \|A\|_p \|B\|_p.   
    \end{align*}
    If, in addition,~$A$ and~$B$ are square then 
    \begin{align*}
        \mu_p(A \oplus B) &= \mu_p(A) + \mu_p(B).
    \end{align*}
\end{Lemma}

\subsection{Relation between compound matrices and Kronecker products}\label{subsec:kron_mul}
This section reviews  results from the recent paper~\cite{Ofir2024Kronecker}  that allow to express  multiplicative [additive] compound matrices using  Kronecker products [sums]. As we will see below, this is useful  in the analysis of compounds of block matrices.

We begin by defining two matrices that provide
a bridge between multiplicative [additive] compound matrices and Kronecker products [sums].
 Fix~$n > 0$ and~$k \in \{1,\dots,n\}$. Let~$e^1,\dots,e^n \in \R^n$ denote the canonical basis of~$\R^n$, and define   matrices~$L_{n,k}\in\R^{ \binom{n}{k}  \times n^k}$ and~$ M_{n,k}\in\R^{n^k \times \binom{n}{k} } $ by
\begin{align}\label{eq:defmn}
    L_{n,k} &:= \sum_{1 \le i_1 < \dots < i_k \le n} \begin{bmatrix} e^{i_1} & \cdots & e^{i_k} \end{bmatrix}^{(k)} (e^{i_1} \otimes \dots \otimes e^{i_k})^T, \nonumber\\
    M_{n,k} &:= \sum_{1 \le i_1 < \dots < i_k \le n} \sum_{(j_1,\dots,j_k) \in S(i_1,\dots,i_k)} \sigma(j_1,\dots,j_k) \nonumber \\&\times(e^{j_1} \otimes \dots \otimes e^{j_k}) \left(\begin{bmatrix} e^{i_1} & \cdots & e^{i_k} \end{bmatrix}^{(k)}\right)^T,
\end{align}
where $S(i_1,\dots,i_k)$ denotes the set of all permutations of the sequence $i_1,\dots,i_k$, and $\sigma(\cdot)$ denotes the signature of the given permutation. 
\begin{example}
    Consider the case~$n=m=2$. Then
    \begin{align*} 
    L_{2,2} & = \sum_{1 \le i_1 <    i_2 \le 2} \begin{bmatrix} e^{i_1} &   e^{i_2} \end{bmatrix}^{(2)} (e^{i_1}    \otimes e^{i_2})^T  \\
    &=\begin{bmatrix}
        1&0\\0&1 
    \end{bmatrix} ^{(2)} \left (  
    \begin{bmatrix}
        1\\0
    \end{bmatrix}\otimes \begin{bmatrix}
        0\\1
    \end{bmatrix}
    \right ) ^T\\
&=\begin{bmatrix}
    0& 1 &0& 0
\end{bmatrix},
    \end{align*}
and
 \begin{align*} 
    M_{2,2} & =   \sum_{(j_1,j_2) \in S(1,2)} \sigma(j_1, j_2) (e^{j_1}   \otimes e^{j_2}) \left(\begin{bmatrix} e^{1} &  e^{2} \end{bmatrix}^{(2)}\right)^T\\
    &= \begin{bmatrix}
        1\\0
    \end{bmatrix}\otimes \begin{bmatrix}
        0\\1
    \end{bmatrix} - \begin{bmatrix}
        0\\1
    \end{bmatrix}\otimes \begin{bmatrix}
        1\\0
    \end{bmatrix}  \\
    &=\begin{bmatrix}
        0&1&-1&0
    \end{bmatrix}^T.
\end{align*}

\end{example}

The
matrices in~\eqref{eq:defmn}
satisfy the relation 
\be\label{eq:L_M_IS_I}
L_{n,k}M_{n,k} = I_r, \text{ with } r := \binom{n}{k}.
\ee
For example, for~$n =k=2$, we have~$r=1$ and 
\[
L_{2,2}M_{2,2}= \begin{bmatrix}
    0& 1 & 0& 0
\end{bmatrix} \begin{bmatrix}
        0&1&-1&0
    \end{bmatrix}^T=1  . 
\]

The next two results from~\cite{Ofir2024Kronecker} relate the $k$-multiplicative [$k$-additive] compound to the $k$th Kronecker power [$k$th Kronecker sum].

\begin{Proposition}
\cite{Ofir2024Kronecker}
\label{thm:MNK+and_LNK}
    Fix~$x^1,\dots,x^k \in \R^n$ with $k \le n$. Then  
\begin{align}
& \begin{bmatrix}x^1 & \dots & x^k\end{bmatrix}^{(k)} = L_{n,k} \nonumber \\
&\times \sum_{(j_1,\dots,j_k) \in S(1,\dots,k)} \sigma(j_1,\dots,j_k)  (x^{j_1} \otimes \dots \otimes x^{j_k}) , \label{eq:lnk_general}
\end{align}
and
\begin{align}
  &  M_{n,k} \begin{bmatrix}x^1 & \dots & x^k\end{bmatrix}^{(k)}\nonumber\\
  &= \sum_{(j_1,\dots,j_k) \in S(1,\dots,k)} \sigma(j_1,\dots,j_k) (x^{j_1} \otimes \dots \otimes x^{j_k}) .\label{eq:mnk_general}
\end{align}
\end{Proposition}

In particular, for~$k=2$ this gives
\begin{align}\label{eq:ln2effect}
    \begin{bmatrix}x^1 & x^2\end{bmatrix}^{(2)}& = L_{n,2} (x^1 \otimes x^2 - x^2 \otimes x^1)   , 
    \end{align}
and
\begin{align}\label{eq:mn2effect}
    M_{n,2} \begin{bmatrix}x^1 & x^2\end{bmatrix}^{(2)}& =  x^1 \otimes x^2 - x^2 \otimes x^1   .
\end{align}

\begin{Proposition}\label{prop:mul_add_kron} \cite{Ofir2024Kronecker}
    Let~$A \in \R^{n \times m}$ and fix~$k \in \{1,\dots,\min\{n,m\}\}$. Then
    \begin{align}\label{eq:mnaa}
        M_{n,k} A^{(k)} &= A^{\otimes k} M_{m,k},  \\
        A^{(k)} &= L_{n,k} A^{\otimes k} M_{m,k}.\nonumber
    \end{align}
    Furthermore, if $n=m$, then
    \begin{equation}\label{eq:aknewp}
        A^{[k]} = L_{n,k} A^{\oplus k} M_{n,k}.
    \end{equation}
\end{Proposition}
Note that the second equation in~\eqref{eq:mnaa} follows from left multiplying both sides of the first equation by~$L_{n,k}$ and using~\eqref{eq:L_M_IS_I}.

\begin{example}
    Consider the case~$n=m=2$ and~$k=2$. Then the terms in the first equation of~\eqref{eq:mnaa} become
    \begin{align*}
        M_{2,2}A^{(2)} &= \begin{bmatrix}
        0&1&-1&0
    \end{bmatrix}^T  \det(A) ,
    \end{align*}
and
    \begin{align*}
            A^{\otimes 2} M_{2,2}& =  \begin{bmatrix}
        a_{11}A&a_{12}A\\
        a_{21}A&a_{22} A 
    \end{bmatrix} \begin{bmatrix}
        0&1&-1&0
    \end{bmatrix}^T \\
    &=  \begin{bmatrix}
        0& 
        \det(A) & 
        -\det(A)&
        0
    \end{bmatrix}^T.
\end{align*}
Also, the terms in~\eqref{eq:aknewp} 
become~$A^{[2]}=\tr(A)= a_{11}+a_{22}$, and
\begin{align*}
    L_{2,2}&(A \oplus A)M_{2,2}= \begin{bmatrix}
    0&1&0&0
    \end{bmatrix} 
\left( \begin{bmatrix}
    a_{11}&0&a_{12} &0 \\
    0&a_{11}&0&a_{12}\\
    a_{21} &0&a_{22} &0\\
    0&a_{21}&0&a_{22}
\end{bmatrix} \right . \\
& + \left . 
\begin{bmatrix}
   a_{11}&a_{12}&0&0\\
   a_{21}&a_{22}&0&0\\
   0&0&a_{11}&a_{12}\\
   0&0&a_{21}&a_{22}
\end{bmatrix}\right)
\begin{bmatrix}
        0&1&-1&0
    \end{bmatrix}^T\\
    &= a_{11}+a_{22}. 
\end{align*}
\end{example}

\subsection{Block Kronecker products and sums}
In general, the compounds of a block matrix mix entries from the various blocks and thus lead to expressions that are hard to understand in terms of the original blocks~\cite{Aikten1935NormalForm}.
The next simple example demonstrates  this. 

\begin{example}  
    Consider the block matrix~$\begin{bmatrix}
        A&B\\C&D
    \end{bmatrix}=\begin{bmatrix}
        a_{11}&a_{12} &b_1\\
        a_{21}&a_{22} & b_2\\
        c_1&c_2 & d
    \end{bmatrix}$. 
    Applying the formula in Prop.~\ref{prop:Explicit_exp_A_k} gives
    \[
   \begin{bmatrix}
        A&B\\C&D
    \end{bmatrix}^{[2]}=\begin{bmatrix}  
        a_{11}+a_{22} & b_2& -b_1 \\c_2& a_{11}+d & a_{12}\\
        -c_1 & a_{21} & a_{22}+d
    \end{bmatrix}.
    \]
 \end{example}

The same problem of mixing entries from different blocks appears when studying Kronecker products of block matrices~\cite{Hyland1989block}. 
To overcome this difficulty, we
 use the \emph{block Kronecker product} defined in~\cite{Hyland1989block} by
\begin{equation}\label{eq:def_block_kron}
    \begin{bmatrix}
        A & B \\
        C & D
    \end{bmatrix} \blkotimes
    \begin{bmatrix}
        E & F \\
        G & H
    \end{bmatrix} :=
    \begin{bmatrix}
        A \otimes E & A \otimes F & B \otimes E & B \otimes F \\
        A \otimes G & A \otimes H & B \otimes G & B \otimes H \\
        C \otimes E & C \otimes F & D \otimes E & D \otimes F \\
        C \otimes G & C \otimes H & D \otimes G & D \otimes H
    \end{bmatrix}.
\end{equation}
Using this definition and~\eqref{eq:mixed_prod} implies that
if $A_i,B_i,C_i,D_i,E_i,F_i,G_i,H_i$, $i=1,2$, 
are matrices such that the products 
$\begin{bmatrix}
        A_1 & B_1 \\
        C_1 & D_1
    \end{bmatrix} \begin{bmatrix}
        A_2 & B_2 \\
        C_2 & D_2
    \end{bmatrix}$ and $\begin{bmatrix}
        E_1 & F_1 \\
        G_1 & H_1
    \end{bmatrix} \begin{bmatrix}
        E_2 & F_2 \\
        G_2 & H_2
    \end{bmatrix}$    
  are well-defined then
\begin{align*}
   &\left(  \begin{bmatrix}
        A_1 & B_1 \\
        C_1 & D_1
    \end{bmatrix} \blkotimes
    \begin{bmatrix}
        E_1 & F_1 \\
        G_1 & H_1
    \end{bmatrix}\right)  \left(  \begin{bmatrix}
        A_2 & B_2 \\
        C_2 & D_2
    \end{bmatrix} \blkotimes
    \begin{bmatrix}
        E_2 & F_2 \\
        G_2 & H_2
    \end{bmatrix}\right) \\&=
    \left( \begin{bmatrix}
        A_1 & B_1 \\
        C_1 & D_1
    \end{bmatrix} \begin{bmatrix}
        A_2 & B_2 \\
        C_2 & D_2
    \end{bmatrix}\right)
    \blkotimes
    \left( \begin{bmatrix}
        E_1 & F_1 \\
        G_1 & H_1
    \end{bmatrix} \begin{bmatrix}
        E_2 & F_2 \\
        G_2 & H_2
    \end{bmatrix}\right),
\end{align*}
that is, the block Kronecker product also satisfies a mixed product property. 
\subsection{Contraction with respect to hierarchic norms}\label{subsec:hier_cont}
Our sufficient condition for 2-contraction in feedback systems combines block
Kronecker products 
and a known hierarchic condition for standard contraction~\cite{Strom1975Logarithmic,Russo2013NetworkContractive}. 

Given~$x \in \R^s$,   partition it into~$m$ sub-vectors as
\[ 
    x = \begin{bmatrix}
        x^1 \\
        \vdots \\
        x^m
    \end{bmatrix},  \text{ with } x^1 \in \R^{s_1},\dots, x^m \in \R^{s_m}, \; \sum_{i=1}^m s_i = s.
\]
Given~$B \in \R^{s \times s}$, 
partition it into blocks compatible with  the partition of~$x$:
\begin{equation*}
    B = \begin{bmatrix}
        B^{11} & \cdots & B^{1m} \\
        \vdots & \ddots & \vdots \\
        B^{m1} & \cdots & B^{mm}
    \end{bmatrix},   
\end{equation*}
with~$B^{ij}\in\R^{s_i\times s_j}$. 

Let~$|\cdot|_i:\R^{s_i}\to\R_{\geq 0}$, $i = 1,\dots,m$, be arbitrary vector norms, 
  and define induced matrix
 norms~$\|\cdot\|_{ij}:\R^{ s_i\times s_j  } \to \R_{\geq 0}$ by
 \[
 \|B^{ij}\|_{ij}:= 
 \sup_{z \in \R^{s_j}\setminus\{0\}} \frac{|B^{ij} z|_i}{|z|_j}. 
 \]
Define a  matrix~$C \in \R^{m \times m}$ by
    \begin{equation}\label{eq:def_b_mat}
        c_{ij} := \begin{cases}
            \mu_i(B^{ii}), & i=j, \\
            \|B^{ij}\|_{ij}, & \text{otherwise}.
        \end{cases}
    \end{equation}

Fix a monotonic
vector norm~$|\cdot|_0:\R^m\to\R_{\geq 0}$, and let~$\mu_0:\R^{m\times m} \to \R$ denote the corresponding matrix measure. 
Define a hierarchic
norm~$|\cdot|:\R^s\to\R_{\geq 0}$ by
\[
  |x|_h := \left|\begin{matrix}
            |x^1|_1 \\
            \vdots \\
            |x^m|_m
        \end{matrix} \right|_0, 
\]
and let~$\mu_h:\R^{s\times s} \to \R$ denote the corresponding matrix measure.  It was
shown in~\cite{Strom1975Logarithmic,Russo2013NetworkContractive} that
\be\label{eq:hier_bound}
\mu_h(B)\leq \mu_0(C). 
\ee
Note that this bounds the matrix measure of the~$s\times s$ matrix~$B$
using the matrix measure of the smaller, $m\times m$ matrix~$C$. In particular, if~$\mu_0(C)\leq -\eta <0$ then~$\mu_h(B)\leq -\eta<0$.

By~\eqref{eq:def_b_mat},   the matrix~$C$ is Metzler. Matrix measures induced by monotonic vector norms are monotonic for Metzler matrices, as shown by the next result.
\begin{Lemma}[\cite{OfirBM2022MinimumEffort}]\label{lem:mono_meas}
    Let $A,B \in \R^{n \times n}$ be Metzler and such that~$a_{ij} \le b_{ij}$ for all $i,j \in \{1,\dots,n\}$. If~$\mu(\cdot)$ is a matrix measure induced by a monotonic vector norm then
   $
        \mu(A) \le \mu(B)$.
\end{Lemma}

\subsection{Network small-gain theorem for Metzler matrices}
In this section we review graph theoretic conditions~\cite{Xiaoming2021GraphMetzler}
for an irreducible  Metzler matrix to be Hurwitz. Cycles in the graph associated with the matrix   play a critical role. We refer to~\cite{Xiaoming2021GraphMetzler} for a comprehensive treatment and numerous examples.

We review first a few useful notions adapted to our problem from graph
theory.  Let $\mcG$ be a strongly-connected weighted directed graph 
with $n$ nodes and with
weighted self-loops such that, for any pair of nodes $i,j$, with~$i,j\in\until{n}$, we have
\begin{itemize}
\item $m_{ii}<0$ is the  weight of the self-loop at node $i$,
\item $m_{ij}>0$ is the weight of directed edge $(i,j)$.
\end{itemize}

In other words,  the adjacency matrix $M=\{m_{ij}\}_{i,j=1}^n$ of~$\mcG$ is an irreducible Metzler matrix with   all diagonal entries negative.

For such a weighted digraph~$\mcG$,
\begin{enumerate}    
\item a \emph{simple cycle} is a directed cycle (with at least 2 nodes) in
  which only the first and last vertices are equal. We do not consider
  self-loops as simple cycles;

\item two cycles $\phi$ and $\psi$ are \emph{disjoint}, denoted by $\phi
  \perp \psi$, if they have no node in common; and 

\item for each $i\in\{2,\dots,n\}$, we let $\mcC_i$ denote the set of
  simples cycles passing through only nodes $\{1,\dots,i\}$.
\end{enumerate}
We next define the notion of gain of a cycle and gain of a set of cycles.
Both are rational functions of the entries of $M$ defined as follows:
\begin{enumerate}
\item the \emph{gain} of a simple cycle $\phi=(i_1,i_2,\dots,i_{k},i_1)$ is
  \begin{equation*}
  \gamma_\phi(M) ~=~ \left(\frac{m_{i_1i_2}}{-m_{i_2i_2}}\right)
  \left(\frac{m_{i_2i_3}}{-m_{i_3i_3}}\right)\cdots
  \left(\frac{m_{i_ki_1}}{-m_{i_1i_1}}\right),
  \end{equation*}

\item for $i\in\{2,\dots,n\}$, the gain of the set of cycles $\mcC_i$ is
  \begin{multline*}
    \gain{\mcC_i}(M) ~:=~
  \sum_{\phi\,\in\,\mcC_i} \gain{\phi}(M) ~-~
  \sum_{\substack{\phi,\psi\,\in\,\mcC_i\\{\phi\perp\psi}}} \gain{\phi}(M)
  \gain{\psi}(M) \\
  ~+~
  \sum_{\substack{\phi,\psi,\rho\,\in\,\mcC_i\\{\phi\perp\psi, \;
        \phi\perp\rho, \; \psi\perp\rho}}}
  \gain{\phi}(M)\gain{\psi}(M)\gain{\rho}(M) ~-~ \cdots
  \end{multline*}
\end{enumerate}
Note that the summation in the definition of $\gain{\mcC_i}(M)$ is finite
because there is only a finite number of cycles that are pairwise disjoint.

In summary, given a Metzler matrix $M\in\mathbb{R}^{n\times n}$ with associated
digraph $\mcG$, we have defined a set of rational functions
$\{\gain{\mcC_2}(M), \dots, \gain{\mcC_n}(M)\}$.  We are now ready to state
the main result of this section.
\begin{Theorem}[Network small-gain theorem for Metzler matrices]
  \label{thm:NetSmallGain}    
  The irreducible Metzler matrix $M$ is Hurwitz if and only if
  \begin{equation}
  \gain{\mcC_2}(M)<1, \dots, \gain{\mcC_n}(M)<1.
  \end{equation}
\end{Theorem}

These  provide    at most $n-1$  small gain conditions for Hurwitzness  (some of the conditions may be redundant) using 
polynomial functions of the
entries of~$M$. The conditions are not unique because they depend upon the numbering
of the nodes.

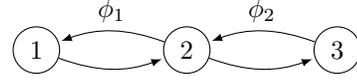
\begin{figure}[t]
    \centering
    \begin{tikzpicture}[scale=1, transform shape]
      \node[state,minimum size = 0.2cm] at (0, 0)     (nodetwo)     {{$1$}};
      \node[state,minimum size = 0.2cm] at (2, 0) (nodeone) {$2$};
      \node[state,minimum size = 0.2cm] at (4, 0)     (nodesix)     {{$3$}};
      \draw[every loop,	auto=right,	>=latex,   ]			
      (nodeone) edge[bend right=20, auto=right] node {$\phi_1$} (nodetwo)
      (nodetwo) edge[bend right=20, auto=right] node {} (nodeone)
      (nodesix) edge[bend right=20, auto=right] node {$\phi_2$} (nodeone)
      (nodeone) edge[bend right=20, auto=right] node {} (nodesix);			
    \end{tikzpicture}
    \caption{Digraph $\mcG$ associated to the Metzler matrix $M$ in
      equation~\eqref{eq:simpleM}.} \label{fig:chain-digraph}
  \end{figure}

We consider a specific example that is required for the treatment
in this paper. Let
\begin{equation} \label{eq:simpleM}
  M = \begin{bmatrix}	
    m_{11} & m_{12} &0\\
    m_{21} & m_{22} & m_{23}\\
    0 & m_{32} & m_{33}
  \end{bmatrix}
\end{equation}	
with~$m_{11},m_{22},m_{33}<0$ and $m_{12}, m_{21},
m_{23}, m_{32}>0$.  We draw the digraph $\mcG$ associated to this matrix $M$ in
Figure~\ref{fig:chain-digraph}. The only simple cycles are~$\phi_1=(1,2,1)$ and~$\phi_2=(2,3,2)$.

For this Metzler matrix, the cycle gains are
\begin{equation}
  \gain{\phi_1}=\frac{m_{12}m_{21}}{m_{11}m_{22}}
  \quad\text{and}\quad
  \gain{\phi_2}=\frac{m_{23}m_{32}}{m_{22}m_{33}};
\end{equation}
the cycle set through nodes $\{1,2\}$ is $\mcC_2=\{\phi_1\}$ so that its
cycle set gain is
\begin{equation}
  \gain{\mcC_2}(M) = \gain{\phi_1}(M); 
\end{equation}
and the cycle set through nodes $\{1,2,3\}$ is $\mcC_3=\{\phi_1,\phi_2\}$ so
that its cycle set gain is
\begin{equation}  
  \gain{\mcC_3}(M) = \gain{\phi_1}(M) + \gain{\phi_2}(M) .
\end{equation}
There are no second (or higher) order terms in
$\gain{\mcC_3}(M)$ because $\phi_1$ and $\phi_2$ are not disjoint.  We also
note that the condition $\gain{\phi_1}<1$ is implied by the condition
$\gain{\phi_1}+\gain{\phi_2}<1$ and is, therefore, redundant.  
Theorem~\ref{thm:NetSmallGain} implies the following result.

\begin{Corollary}\label{cor:chain-matrix}
  The Metzler matrix $M$ in equation~\eqref{eq:simpleM}, with~$m_{ii}<0$,
  is Hurwitz if and
  only if $\ds \gain{\phi_1} + \gain{\phi_2}<1$, that is, iff 
  \begin{equation}\label{condition:chain-matrix}
  \frac{m_{12}m_{21}}{m_{11}m_{22}} + \frac{m_{23}m_{32}}{m_{22}m_{33}}<1.
  \end{equation}
\end{Corollary}

For example, consider the case when $m_{11}=m_{22}=m_{33}=-c<0$ and
$m_{12} = m_{21} = m_{23} = m_{32} = 1$. Then~\eqref{condition:chain-matrix} yields that~$M$ is Hurwitz iff~$c>\sqrt{2}$.
   This
  can be easily verified since the spectrum of
$\begin{bmatrix}\begin{smallmatrix} -c & 1 & 0 \\ 1 & -c & 1 \\ 0 & 1 & -c
\end{smallmatrix}\end{bmatrix}$ is $\{-c, -c-\sqrt{2},-c+\sqrt{2}\}$.

\section{Main result}\label{sec:main}
Consider a general feedback interconnection of two dynamical sub-systems given by
\begin{equation}\label{eq:feedback}
    \begin{aligned}
        \dot x &= f(x,z), \\
        \dot z &= g(x,z),
    \end{aligned}
\end{equation}
with~$x : [0,\infty) \to \Omega_x \subseteq \R^n, z: [0,\infty) \to \Omega_z \subseteq \R^m$, and $f,g$   continuously differentiable in both parameters. We assume throughout
that the state-space~$\Omega_x \times \Omega_z$ is convex,
and that~$n,m\geq 2$. 

\begin{figure*}[!t] 
\normalsize
\[
S(x,z):= \begin{bmatrix} \mu_{p,T_A^{(2)}}\mleft(\frac{\partial f}{\partial x}^{[2]}(x,z)\mright) & 2^{1/q}\mleft\|T_A \frac{\partial f}{\partial z}(x,z) T_D^{-1}\mright\|_{p} & 0 \\
            2^{1/p} \mleft\| T_D \frac{\partial g}{\partial x}(x,z) T_A^{-1} \mright\|_{p} & \mu_{p,T_A}\mleft(\frac{\partial f}{\partial x}(x,z)\mright) + \mu_{p,T_D}\mleft(\frac{\partial g}{\partial z}(x,z)\mright) & 2^{1/p} \mleft\| T_A \frac{\partial f}{\partial z}(x,z) T_D^{-1} \mright\|_{p} \\
            0 & 2^{1/q} \mleft\| T_D \frac{\partial g}{\partial x}(x,z) T_A^{-1} \mright\|_{p} & \mu_{p,T_D^{(2)}}\mleft(\frac{\partial g}{\partial z}^{[2]}(x,z)\mright)
        \end{bmatrix}.
\]
\caption{Definition of the matrix~$S(x,z)$ in Thm.~\ref{thm:2cont_feedback}.  \label{fig:def_S}}
\hrulefill
\vspace*{4pt}
\end{figure*}

We can now state our main result. This provides a sufficient condition for 2-contraction
of the feedback system~\eqref{eq:feedback} that is based on the matrix measure of a~$3\times 3$ Metzler  matrix.
Recall that if~$\mu :\R^{n\times n}\to\R$ is the matrix measure corresponding to  a  norm~$|\cdot| :\R^n\to \R_{\geq 0}$, and~$T \in\R^{n\times n}$ is non-singular then the matrix measure corresponding  to the scaled norm~$|z|_T:=|Tz|$ is~$\mu_T(A)= \mu (TAT^{-1})$. 
For an~$L_p$ norm~$|\cdot|_p$, and a non-singular matrix~$T$,
let~$\mu_{p,T}(A) : =\mu_p(TAT^{-1})$
\begin{Theorem}\label{thm:2cont_feedback}
Fix~$p\in[1,\infty]$, and non-singular matrices~$T_A\in\R^{n\times n}$, $T_D\in\R^{m\times m}$. Let~$q$ be such that~$p^{-1}+q^{-1}=1$ (with the convention~$1/\infty=0$). Define~$S: \Omega_x\times \Omega_z \to \R^{3 \times 3} $ as in Fig.~\ref{fig:def_S}. 
If there exists a matrix measure~$\mu_0$ induced by a 
monotonic vector norm~$|\cdot|_0$, such that
\begin{equation}\label{eq:main_cond}
  \mu_0(S(x,z)) \le -\eta < 0 \text{ for all }(x,z) \in \Omega_x \times \Omega_z 
    \end{equation}
      then the feedback system~\eqref{eq:feedback} is 2-contracting with rate~$\eta$.
\end{Theorem}

\begin{Remark}
    To explain this condition, consider for simplicity the case where~$T_A=I_n$
    and~$T_D=I_m$. Then the condition  in~\eqref{eq:main_cond} becomes
\[
    \mu_0\mleft( \begin{bmatrix}
            \mu_{p }\mleft(\frac{\partial f}{\partial x}^{[2]} \mright) & 2^{1/q}\mleft\|  \frac{\partial f}{\partial z}   \mright\|_{p} & 0 \\
            2^{1/p} \mleft\|   \frac{\partial g}{\partial x}    \mright\|_{p} & \mu_{p }\mleft(\frac{\partial f}{\partial x} \mright) + \mu_{p }\mleft(\frac{\partial g}{\partial z}  \mright) & 2^{1/p} \mleft\|  \frac{\partial f}{\partial z}   \mright\|_{p} \\
            0 & 2 ^{1/q} \mleft\|  \frac{\partial g}{\partial x}    \mright\|_{p} & \mu_{p }\mleft(\frac{\partial g}{\partial z}^{[2]} \mright) \end{bmatrix}\mright) \le -\eta  .
\]  
If this holds then in particular all the diagonal terms in the matrix must be negative, that is,~$\dot x=f(x,z)$ is 2-contractive for any~$z\in\Omega_z$, $\dot z=g(z,x)$ is 2-contractive for any~$x\in\Omega_x$,
and~$
 \mu_{p }\mleft(\frac{\partial f}{\partial x}(x,z)\mright) + \mu_{p }\mleft(\frac{\partial g}{\partial z}(x,z)\mright)<0$
for all~$(x,z) \in \Omega_x \times \Omega_z$. Note that this last condition is a kind of ``additive  1-contraction condition'', but it does not require any of the sub-systems to be~$1$-contractive on the entire state-space. Moreover, the diagonal entries
in the matrix
must be ``negative enough'' to compensate  for the off-diagonal entries that represent gains of the ``cross terms'' $\frac{\partial f}{\partial z}(x,z)$ 
and~$\frac{\partial g}{\partial x}(x,z)$. Summarizing,  this provides  a sufficient
condition for $2$-contraction of the $(n+m)$-dimensional feedback system~\eqref{eq:feedback}    using properties of the two sub-systems and their interconnections. 
\end{Remark}

\subsection{Small-gain theorem for 2-contraction}
We now extend Theorem~\ref{thm:2cont_feedback} giving a condition that is
more conservative, but easier to verify. The key standard idea is that
\begin{equation*}
  \sup_{x,z} \mu_0(S(x,z)) \leq    \mu_0( \sup_{x,z} S(x,z)) ,
\end{equation*}
because of the monotonicity properties of matrix measures induced by monotonic vectors norms (see Lemma~\ref{lem:mono_meas}).

\newcommand{\Lip}{\operatorname{Lip}}
\newcommand{\osL}{\operatorname{osL}}
\renewcommand{\Lip}{\operatorname{\ell}}
\renewcommand{\osL}{\operatorname{c}}

\begin{Corollary}\label{coro:smallgain_2cont}
    Fix~$p \in [1,\infty]$, and non-singular matrices~$T_A \in \R^{n \times n}, T_A \in \R^{m \times m}$. Define further the following quantities (including Lipschitz constants, one-sided Lipschitz constants, and loosely speaking, compounded one-sided Lipschitz constants):
\begin{itemize}[nosep]
\item $\Lip_z(f) := \sup_{x,z} \mleft\|T_A \frac{\partial f}{\partial z}(x,z) T_D^{-1}\mright\|_{p}$, 
\item $\Lip_x(g) := \sup_{x,z} \mleft\| T_D \frac{\partial g}{\partial x}(x,z) T_A^{-1} \mright\|_{p}$,
\item $\osL_x(f) := \sup_{x,z} \mu_{p,T_A}\mleft(\frac{\partial f}{\partial x}(x,z)\mright)$, 
\item $\osL_z(g) := \sup_{x,z} \mu_{p,T_D}\mleft(\frac{\partial g}{\partial z}(x,z)\mright)$,
\item $\osL_x^{[2]}(f) := \sup_{x,z} \mu_{p,T_A^{(2)}}\mleft(\frac{\partial f}{\partial x}^{[2]}(x,z)\mright)$, and
\item $\osL_z^{[2]}(g) := \sup_{x,z} \mu_{p,T_D^{(2)}}\mleft(\frac{\partial g}{\partial z}^{[2]}(x,z)\mright)$.
\end{itemize}
    If the following four conditions hold
    \begin{gather}
      \osL^{[2]}_x(f)<0, \quad \osL_x(f) + \osL_z(g)<0, \quad  \osL_z^{[2]}(g)<0,
      \label{cond:contractivity}
    \\
      \Lip_z(f) \Lip_x(g) < \frac{1}{2}
      \frac{ (\osL_x(f) + \osL_z(g)) \osL^{[2]}_x(f) \osL_z^{[2]}(g)}
      { \osL^{[2]}_x(f)+\osL_z^{[2]}(g)},
      \label{cond:small-gain}
    \end{gather}
    then the feedback system~\eqref{eq:nonlin} is 2-contracting.
\end{Corollary}
\begin{IEEEproof}
Let~$q$ be such that~$p^{-1}+q^{-1}=1$.
Define the constant Metzler matrix
    \begin{equation*}
        S_{\max} := \begin{bmatrix}
            \osL^{[2]}_x(f) & 2^{1/q}\Lip_z(f) & 0 \\
            2^{1/p}\Lip_x(g) & \osL_x(f) + \osL_z(g) & 2^{1/p}\Lip_z(f) \\
            0 & 2^{1/q} \Lip_x(g) & \osL^{[2]}_z(g)
        \end{bmatrix}.
    \end{equation*}
 Condition~\eqref{cond:contractivity}     implies that~$S_{\max}$ is Metzler with negative diagonal entries.     Corollary~\ref{cor:chain-matrix}, and the fact that~$2^{1/p}2^{1/q}=2$, implies that  condition~\eqref{cond:small-gain}  guarantees that~$S_{\max}$ is Hurwitz. Thus, there exists 
 a  monotonic vector norm~$|\cdot|_0$ such that~$\mu_0(S_{\max}) < 0$~\cite{Pastravanu2006GenDiagStab}. 
    It is clear that~$S(x,z) \le S_{\max}$,
    so  Lemma~\ref{lem:mono_meas} gives~$\mu_0(S(x,z) ) \le \mu_0(S_{\max})<0$ for all~$x,z$, and using Theorem~\ref{thm:2cont_feedback} 
      completes the proof.
\end{IEEEproof}

For context, the standard small-gain 1-contractivity theorem states
that the interconnection is 1-contracting (i.e., contracting)  if 
\begin{gather}
  \osL_x(f)<0, \quad \osL_z(g)<0,
  \label{cond:contractivity:1}
  \\
   \Lip_z(f) \Lip_x(g) <  \osL_x(f) \osL_z(g) , 
     \label{cond:small-gain:1}
\end{gather}
(see, e.g.~\cite[Section~3.6]{Bullo2024ContractionBook}).
 
\section{Proof of main result}\label{sec:proof}
The proof of Thm.~\ref{thm:2cont_feedback}  includes two main steps. The first is deriving   explicit expressions for the 2-compounds of a block matrix. The second is deriving an upper bound on the matrix measure of the 2-additive compound.

\subsection{Explicit expression for the 2-compounds of a block matrix}
The first step in the proof of our main result is based on deriving a novel   explicit expression for the 2-multiplicative   compound of a block matrix
$    \begin{bmatrix}
        A & B \\
        C & D
    \end{bmatrix}$,
    with~$A\in\R^{n\times n}$,
    $B\in\R^{n\times m}$, $C\in\R^{m\times n}$, and~$D\in\R^{m\times m}$.
We believe that this expression is of independent interest, and will find more applications in the analysis of   of feedback  systems. 

  We then use this   to derive
  an expression for the $2$-additive compound of a block matrix. which is simpler than
  the expression    derived in  a    recent   paper by Angeli et al.~\cite{Angeli2023smallgain}. Our approach 
is different from the one used  in~\cite{Angeli2023smallgain}, and we believe that it 
can be 
extended to arbitrary orders of compounds. 

We require   several auxiliary results.
The  first  result provides an expression for
the~$2$-multiplicative compound of a 
special ``vector block matrix'' in terms of its blocks.
\begin{Lemma}\label{lem:block_mul}
    Fix $x^1,x^2 \in \R^n$ and $y^1,y^2 \in \R^m$,  with $n,m \ge 2$. There exists a permutation matrix~$P_{n,m} \in \R^{r \times r}$, with~$r := \binom{n+m}{2} =\binom{n}{2}+nm+\binom{m}{2} $, such that
    \begin{equation}\label{eq:x1x2}
        P_{n,m} \begin{bmatrix}
            x^1 & x^2 \\
            y^1 & y^2
        \end{bmatrix} ^{(2)}= \begin{bmatrix}
            \begin{bmatrix} x^1 & x^2 \end{bmatrix}^{(2)} \\
            x^1 \otimes y^2 - x^2 \otimes y^1 \\
            \begin{bmatrix} y^1 & y^2 \end{bmatrix}^{(2)}
        \end{bmatrix}.
    \end{equation}
\end{Lemma}
\begin{IEEEproof}
    Let $  
        Q := \begin{bmatrix}
            x^1 & x^2 \\
            y^1 & y^2
        \end{bmatrix}\in\R^{ (n+m)\times 2 }$.
    The entries of~$Q^{(2)}$ are all the~$2\times 2$
    minors of~$Q$, which can be divided into three groups: (1) minors of submatrices whose rows belong to the first~$n$ rows of~$Q$; (2) minors of submatrices whose rows belong to the last~$m$ rows of~$Q$; and (3) minors of submatrices where the first row belongs to the first~$n$ rows of~$Q$ and the second row belongs to the last~$m$ rows of~$Q$. The minors of 
    groups~(1) and~(2) are exactly all the entries of~$\begin{bmatrix} x^1 & x^2 \end{bmatrix}^{(2)}$ and~$\begin{bmatrix} y^1 & y^2 \end{bmatrix}^{(2)}$, respectively. The minors of group~(3) are all terms~$x^1_i y^2_j - x^2_i y^1_j$, with $i \in \{1,\dots,n\}$ and~$j \in \{1,\dots,m\}$.
    By definition of the Kronecker product, these are  all the entries of the vector~$x^1 \otimes y^2 - x^2 \otimes y^1$. 
This completes the proof.
\end{IEEEproof}

\begin{example}
    Consider the case~$n=m=2$.
    To simplify the notation, denote the entries of the vectors by~$x^1=\begin{bmatrix}
        a&b
    \end{bmatrix}^T$,
    $x^2=\begin{bmatrix}
        c&d
    \end{bmatrix}^T$,
    $y^1=\begin{bmatrix}
        e&f
    \end{bmatrix}^T$, and~$y^2=\begin{bmatrix}
        g&h
    \end{bmatrix}^T$.  
Then
\begin{align*}
\begin{bmatrix}
            x^1 & x^2 \\
            y^1 & y^2
        \end{bmatrix} ^{(2)}= \begin{bmatrix}
            ad-bc \\ag-ce\\ah-cf\\bg-de\\bh-df\\eh-gf
        \end{bmatrix},
\end{align*}
whereas
\begin{align*}
    \begin{bmatrix}
        \begin{bmatrix} x^1 & x^2 \end{bmatrix}^{(2)} \\
            x^1 \otimes y^2 - x^2 \otimes y^1 \\
            \begin{bmatrix} y^1 & y^2 \end{bmatrix}^{(2)}
        \end{bmatrix}=\begin{bmatrix}
            ad-bc\\
            ag-ce\\ah-cf\\bg-de\\bh-df
            \\
            eh-fg
        \end{bmatrix}.
\end{align*}
Thus,~\eqref{eq:x1x2}  holds with~$P_{2,2}=I_6$.
\end{example}

The next result allows to rewrite the expression on the right-hand side of~\eqref{eq:x1x2}
as a block Kronecker product. 

\begin{Lemma}\label{lem:blk_mul_kron_vec}
    Let $x^1,x^2 \in \R^n$ and $y^1,y^2 \in \R^m$,  with $n,m \ge 2$. Let~$P_{n,m} \in \R^{r \times r}$, with $r := \binom{n+m}{2}$, be the permutation matrix in  Lemma~\ref{lem:block_mul}.  
 Define~$Z\in\R^{r\times (n+m)^2}$ and~$V\in\R^{   (n+m)^2 \times r }$   
by
\be\label{eq:def_mat_Z}
Z:=
\begin{bmatrix}
            L_{n,2} & 0 & 0 & 0 \\
            0 & I_{nm} & 0 & 0 \\
            0 & 0 & 0 & L_{m,2}
        \end{bmatrix}
,
\ee
\[
V:=\begin{bmatrix}
            M_{n,2} & 0 & 0 \\
            0 & I_{n m} & 0 \\
            0 & -H_{m,n} & 0 \\
            0 & 0 & M_{m,2}
        \end{bmatrix}.
\]
    Then
\begin{equation}\label{eq:block_mul_kron_Mn}
      V P_{n,m} \begin{bmatrix}
            x^1 & x^2 \\
            y^1 & y^2
        \end{bmatrix}^{(2)} = 
        \begin{bmatrix}
            x^1 \\
            y^1
        \end{bmatrix} \blkotimes
        \begin{bmatrix}
            x^2 \\
            y^2
        \end{bmatrix} - \begin{bmatrix}
            x^2 \\
            y^2
        \end{bmatrix} \blkotimes
        \begin{bmatrix}
            x^1 \\
            y^1
        \end{bmatrix}, 
    \end{equation}
    and    \begin{equation}\label{eq:block_mul_kron_Ln}
        P_{n,m} \begin{bmatrix}
            x^1 & x^2 \\
            y^1 & y^2
        \end{bmatrix}^{(2)} = Z \left(
        \begin{bmatrix}
            x^1 \\
            y^1
        \end{bmatrix} \blkotimes
        \begin{bmatrix}
            x^2 \\
            y^2
        \end{bmatrix} - \begin{bmatrix}
            x^2 \\
            y^2
        \end{bmatrix} \blkotimes
        \begin{bmatrix}
            x^1 \\
            y^1
        \end{bmatrix}
        \right). 
    \end{equation}
\end{Lemma}

\begin{IEEEproof}
By Lemma~\ref{lem:block_mul}, 
 \begin{align*}
     V   P_{n,m} \begin{bmatrix}
            x^1 & x^2 \\
            y^1 & y^2
        \end{bmatrix} ^{(2)}&= V \begin{bmatrix}
            \begin{bmatrix} x^1 & x^2 \end{bmatrix}^{(2)} \\
            x^1 \otimes y^2 - x^2 \otimes y^1 \\
            \begin{bmatrix} y^1 & y^2 \end{bmatrix}^{(2)}
        \end{bmatrix}\\
        &= \begin{bmatrix}
        M_{n,2}  \begin{bmatrix} x^1 & x^2 \end{bmatrix}^{(2)} \\
         x^1 \otimes y^2 - x^2 \otimes y^1 \\
         -H_{m,n} ( x^1 \otimes y^2 - x^2 \otimes y^1) \\
         M_{m,2} \begin{bmatrix} y^1 & y^2 \end{bmatrix}^{(2)}
        \end{bmatrix},
 \end{align*}
 and using~\eqref{eq:mn2effect}    and~\eqref{eq:a_otimes_b} gives
  \begin{align*}
     V   P_{n,m} \begin{bmatrix}
            x^1 & x^2 \\
            y^1 & y^2
        \end{bmatrix} ^{(2)}&=
\begin{bmatrix}
        x^1 \otimes x^2 - x^2 \otimes x^1 \\
        x^1 \otimes y^2 - x^2 \otimes y^1 \\
        y^1 \otimes x^2 - y^2 \otimes x^1 \\
        y^1 \otimes y^2 - y^2 \otimes y^1
    \end{bmatrix}.
\end{align*}
 Using  the definition of the block Kronecker product~$\blkotimes$ in~\eqref{eq:def_block_kron} proves~\eqref{eq:block_mul_kron_Mn}. Now multiplying both sides of~\eqref{eq:block_mul_kron_Mn} by $Z$ from the left,  
 and using the fact that
\be\label{eq:ZVProd} 
ZV =I_r, 
    \text{ with } r := \binom{n+m}{2},
    \ee
which follows from~\eqref{eq:L_M_IS_I},
proves~\eqref{eq:block_mul_kron_Ln}. This completes the proof of  Lemma~\ref{lem:blk_mul_kron_vec}.
\end{IEEEproof}

The next result relates the $2$-multiplicative compound of a general block matrix to a block Kronecker product. 
\begin{Proposition}\label{prop:gene}
    Fix integers~$n,m \ge 2$, and let~$P_{n,m}$ be as in Lemma~\ref{lem:block_mul}. Then for any~$A \in \R^{n \times n}, B \in \R^{n \times m}, C \in \R^{m \times n}$, and~$D \in \R^{m \times m}$, we have 
    \begin{equation}\label{eq:mul_blk_implicit}
        P_{n,m}
        \begin{bmatrix}
            A & B \\
            C & D
        \end{bmatrix}^{(2)}
        P_{n,m}^T = 
        Z
        \left(
        \begin{bmatrix}
            A & B \\
            C & D
        \end{bmatrix}
        \blkotimes
        \begin{bmatrix}
            A & B \\
            C & D
        \end{bmatrix}
        \right)
        V. 
    \end{equation}
\end{Proposition}
\begin{IEEEproof}
Pick arbitrary vectors~$x^1,x^2 \in \R^n$ and~$y^1,y^2 \in \R^m$, and define~$q\in \R^r$ by
\be\label{eq:def_mat_QQ}
q :=  \left(
        \begin{bmatrix}
            A & B \\
            C & D
        \end{bmatrix}
        \blkotimes
        \begin{bmatrix}
            A & B \\
            C & D
        \end{bmatrix}
        \right)
        V
        P_{n,m}
        \begin{bmatrix}
            x^1 & x^2 \\
            y^1 & y^2
        \end{bmatrix}^{(2)}.
\ee
Then by~\eqref{eq:block_mul_kron_Mn} and the mixed product property,
    \begin{align*}
   q
        &= 
        \left(
        \begin{bmatrix}
            A & B \\
            C & D
        \end{bmatrix}
        \blkotimes
        \begin{bmatrix}
            A & B \\
            C & D
        \end{bmatrix}
        \right)
        \left(
        \begin{bmatrix}
            x^1 \\
            y^1 
        \end{bmatrix}
        \blkotimes
        \begin{bmatrix}
            x^2 \\
            y^2
        \end{bmatrix} - 
        \begin{bmatrix}
            x^2 \\
            y^2 
        \end{bmatrix}
        \blkotimes
        \begin{bmatrix}
            x^1 \\
            y^1
        \end{bmatrix}
        \right)
        \\
        &=
        \begin{bmatrix}
            Ax^1 + By^1 \\
            Cx^1 + Dy^1
        \end{bmatrix}
        \blkotimes
        \begin{bmatrix}
            Ax^2 + By^2 \\
            Cx^2 + Dy^2
        \end{bmatrix} \\&- 
        \begin{bmatrix}
            Ax^2 + By^2 \\
            Cx^2 + Dy^2 
        \end{bmatrix}
        \blkotimes
        \begin{bmatrix}
            Ax^1 + By^1 \\
            Cx^1 + Dy^1
        \end{bmatrix} .
        \end{align*}
        Using~\eqref{eq:block_mul_kron_Mn} again gives
        $
        q=
        V P_{n,m} \begin{bmatrix}
            Ax^1 + By^1 & Ax^2 + By^2 \\
            Cx^1 + Dy^1 & Cx^2 + Dy^2
        \end{bmatrix}^{(2)}
        $, and applying the  Cauchy-Binet Theorem gives
        \begin{align*} 
        q&=
        V P_{n,m} \begin{bmatrix}
            A & B \\
            C & D
        \end{bmatrix}^{(2)} \begin{bmatrix}
            x^1 & x^2 \\
            y^1 & y^2
        \end{bmatrix}^{(2)}.
    \end{align*}
  Since the   vectors $x^i,y^i$ are arbitrary, combining this with~\eqref{eq:def_mat_QQ}
    proves that
\begin{equation}\label{eq:block_mul_kron_Mn_mat}
       V
        P_{n,m}
        \begin{bmatrix}
            A & B \\
            C & D
        \end{bmatrix}^{(2)} = 
        \left(
        \begin{bmatrix}
            A & B \\
            C & D
        \end{bmatrix}
        \blkotimes
        \begin{bmatrix}
            A & B \\
            C & D
        \end{bmatrix}
        \right)
       V P_{n,m}.
    \end{equation}
Multiplying this equation on   the right-hand side [left-hand side]  by~$P_{n,m}^T$
[$Z$] gives
\[
     Z  V
        P_{n,m}
        \begin{bmatrix}
            A & B \\
            C & D
        \end{bmatrix}^{(2)} P_{n,m}^T = Z
        \left(
        \begin{bmatrix}
            A & B \\
            C & D
        \end{bmatrix}
        \blkotimes
        \begin{bmatrix}
            A & B \\
            C & D
        \end{bmatrix}
        \right)
       V  , 
\]
and using~\eqref{eq:ZVProd}   completes the proof of Prop.~\ref{prop:gene}.
\end{IEEEproof}

\begin{figure*}[!t] 
\normalsize
  \begin{minipage}[t]{1\textwidth}
  \centering
    \[    
P_{n,m}
        \begin{bmatrix}
            A & B \\
            C & D
        \end{bmatrix}^{(2)}
        P_{n,m}^T = 
       \begin{bmatrix}
        A^{(2)} & M^T_{n,2}(A \otimes B) & B^{(2)} \\
        (A \otimes C)M_{n,2} & A \otimes D - (B \otimes C)H_{m,n} & (B \otimes D)M_{m,2} \\
        C^{(2)} & M^T_{m,2}(C \otimes D) & D^{(2)}
    \end{bmatrix}
    .
\]
\caption{The $2$-multiplicative compound of a block matrix.\label{fig:P_block_P}}\hrulefill
\end{minipage}\\
  \begin{minipage}[t]{1\textwidth}
  \centering
    \[    
        P_{n,m}
        \begin{bmatrix}
            A & B \\
            C & D
        \end{bmatrix}^{[2]}
        P_{n,m}^T =\begin{bmatrix}
            A^{[2]} & M^T_{n,2}(I_n \otimes B) & 0 \\
            (I_n \otimes C)M_{n,2} & A \oplus D & (B \otimes I_m)M_{m,2} \\
            0 & M^T_{m,2}(C \otimes I_m) & D^{[2]} 
        \end{bmatrix}
    .
\]
\caption{The 2-additive compound of a block matrix.\label{fig:2add}}\hrulefill
\end{minipage}\\
\end{figure*}

The next result provides an explicit expression for the $2$-multiplicative compound  of a general block matrix. 
\begin{Theorem}
        \label{thm:block_comp_explicit}
    Fix integers~$n,m \ge 2$ and let~$P_{n,m}$ be as in Lemma~\ref{lem:block_mul}. Then for any~$A \in \R^{n \times n}, B \in \R^{n \times m}, C \in \R^{m \times n}, D \in \R^{m \times m}$, the $2$-multiplicative compound of the block matrix~$\begin{bmatrix}
        A & B \\ C & D 
    \end{bmatrix}$
    satisfies the expression in Fig.~\ref{fig:P_block_P}.
\end{Theorem}
\begin{IEEEproof}
Let~$ X:=P_{n,m}
        \begin{bmatrix}
            A & B \\
            C & D
        \end{bmatrix}^{(2)}
        P_{n,m}^T$.     
Performing the matrix
    multiplications on the right-hand side of~\eqref{eq:mul_blk_implicit},
   and applying  Prop.~\ref{prop:mul_add_kron} gives
   \[
   X= \left[ \begin{smallmatrix}
            A^{(2)} & L_{n,2} (A \otimes B) - L_{n,2} (B \otimes A) H_{m,n} & B^{(2)} \\
            (A \otimes C) M_{n,2} & A \otimes D - (B \otimes C) H_{m,n} & (B \otimes D) M_{m,2} \\
            C^{(2)} & L_{m,2} (C \otimes D) - L_{m,2} (D \otimes C) H_{m,n} & D^{(2)}
        \end{smallmatrix}\right].
   \]
   To simplify this expression note that
 using~\eqref{eq:defmn}
    and~~\eqref{eq:a_otimes_b}
gives
\begin{align*}
        M^T_{n,2} &= \sum_{1 \le i_1 < i_2 \le n} \left ( \begin{bmatrix}
            e^{i_1} & e^{i_2}
        \end{bmatrix}^{(2)} (e^{i_1} \otimes e^{i_2})^T \right .\\
        & \;\;\;\;\;\;\;\; - \left .\begin{bmatrix}
            e^{i_1} & e^{i_2}
        \end{bmatrix}^{(2)} (e^{i_2} \otimes e^{i_1})^T\right) \\
        &= L_{n,2} - L_{n,2} H_{n,n}^T\\
        &= L_{n,2} (I_{n^2} - H_{n,n}).
    \end{align*}
    Therefore, for any $A \in \R^{n \times n}$ and~$ B \in \R^{n \times m}$, we have 
    \begin{align}\label{eq:first_X_simp}
        M^T_{n,2} (A \otimes B) &= L_{n,2} (I_{n^2} - H_{n,n}) (A \otimes B)\nonumber \\
        &= L_{n,2} (A \otimes B) - L_{n,2} (B \otimes A) H_{n,m}^T \nonumber\\
        &= L_{n,2} (A \otimes B) - L_{n,2} (B \otimes A) H_{m,n},
    \end{align}
    where the last equation follows from~\eqref{eq:A_MUL_B_REP}. A similar argument shows that
    \be\label{eq:sec_X_simp}
    M_{m,2}^T
    (C\otimes D)=
    L_{m,2} (C \otimes D) - L_{m,2} (D \otimes C) H_{m,n}. 
    \ee
    Using~\eqref{eq:first_X_simp} and~\eqref{eq:sec_X_simp} to simplify~$X$
     completes the proof. 
\end{IEEEproof}

\begin{Remark}
Fix~$\varepsilon\geq 0$,  and let
\begin{align*}
L(\varepsilon
) & :=P_{n,m} \left (I_{n+m}+\varepsilon
\begin{bmatrix}
    A & B \\ C & D
\end{bmatrix}\right )^{(2)} P_{n,m}^T\\
&= P_{n,m} 
\begin{bmatrix}
   I_n+\varepsilon A & \varepsilon B \\
   \varepsilon C & I_m+\varepsilon D
\end{bmatrix} ^{(2)} P_{n,m}^T.
\end{align*}
Then
\[
         P_{n,m}
        \begin{bmatrix}
            A & B \\
            C & D
        \end{bmatrix}^{[2]} P_{n,m}^T
        = \frac{d}{d\varepsilon} L(\varepsilon) |_{\varepsilon=0}.
\]
Applying  Theorem~\ref{thm:block_comp_explicit}  yields the expression for the $2$-additive compound 
 of a block matrix 
 in Fig.~\ref{fig:2add}.
Angeli et al.~\cite{Angeli2023smallgain} recently derived   a more complicated but equivalent 
expression   
using a 
different approach.
\end{Remark}

\begin{example}
    Consider the  block-diagonal case, that is, the case where in the block matrix we have~$B=0$ and~$C=0$. Then the expressions above simplify to  
    \begin{align*}
        P_{n,m}
        \begin{bmatrix}
            A & 0 \\
            0 & D
        \end{bmatrix}^{(2)}
        P_{n,m}^T &= \begin{bmatrix}
            A^{(2)} & 0 & 0 \\
            0 & A \otimes D & 0 \\
            0 & 0 & D^{(2)}
        \end{bmatrix},  \\
        P_{n,m}
        \begin{bmatrix}
            A & 0 \\
            0 & D
        \end{bmatrix}^{[2]}
        P_{n,m}^T &= \begin{bmatrix}
            A^{[2]} & 0 & 0 \\
            0 & A \oplus D & 0 \\
            0 & 0 & D^{[2]}
        \end{bmatrix},
    \end{align*}
    and this recovers      the expressions in~\cite[Thm.~3]{OfirMLS2021SeriesTAC} when~$k=2$.
\end{example}

Recall that the LTI system~$\dot x = Ax$ is called~$k$-positive if~$A^{[k]}$ is Metzler~\cite{Weiss2021kpositive}, and strongly $k$-positive if, in addition,~$A^{[k]}$ is irreducible. $k$-positivity has been used to generalize  non-linear cooperative systems to~$k$-cooperative systems, and this has found several  applications, see e.g.~\cite{Alseidi2019Discretekpos,Katz2020EntrainOscillatory}. For example, strongly 2-cooperative systems admit a strong Poincar\'e-Bendixson property: any bounded solution that 
keeps a positive distance from the set of equilibria converges to a periodic orbit~\cite{rami_3D,Weiss2021kpositive}.
 As an additional example, we use our explicit expression for the 2-additive compound a block matrix in Fig.~\ref{fig:2add} 
    to find conditions guaranteeing
    that the  interconnection of two   LTI sub-systems is 2-positive.
\begin{Corollary}
Consider system~\eqref{eq:nonlin} and assume both sub-systems are LTI, i.e.,
    \begin{align*}
        f(x,z) &= Ax + Bz, \\
        g(x,z) &= Cx + Dz,
    \end{align*}
    with~$A \in \R^{n \times n}, B \in \R^{n \times m}, C \in \R^{m \times n}$, and~$D \in \R^{m \times m}$.
The feedback system is 2-positive iff all the following two conditions hold:
\begin{enumerate}
    \item The matrices~$A^{[2]}$, $D^{[2]}$, and~$A\oplus D$ are Metzler; and
    \item The matrices   
    $ M^T_{n,2}(I_n \otimes B) $, 
     $       (I_n \otimes C)M_{n,2} $,
     $ (B \otimes I_m)M_{m,2} $,  
            and~$ M^T_{m,2}(C \otimes I_m)$ are  
            non-negative. 
\end{enumerate}
\end{Corollary}
\begin{IEEEproof}
    The proof follows from the expression in Fig.~\ref{fig:2add}, and the fact that the Metzler property is invariant under a coordinate transformation with a permutation matrix. 
\end{IEEEproof}

The expressions for the 2-compounds of a block matrix in
Fig.~\ref{fig:P_block_P} and Fig.~\ref{fig:2add} mix  
the four sub-matrices and the matrices~$L_{n,2}$, $L_{m,2}$, $M_{n,2}$,
$M_{m,2}$, and~$H_{m,n}$. Deriving
sufficient conditions
for contraction requires bounding the norms of the 
matrices
in these expressions. 

\subsection{Bounds on the norms of the blocks}
We now apply
the expressions for the $2$-compounds to   derive sufficient conditions for 2-contraction in a feedback interconnection. 
  We begin by analyzing  the induced matrix $p$-norms 
  of~$ L_{n,k}$ and~$M_{n,k} $, with~$p\in[1,\infty]$. Recall that we  use the convention~$1 / \infty = 0$.
\begin{Lemma}\label{lem:LnkMnk_norms}
    Fix~$k \in \{1,\dots,n\}$ and~$p \in [1,\infty]$.
    Then
    \begin{equation}\label{eq:Lnk_norm}
        \|L_{n,k}\|_p = 1,
    \end{equation}
    and
    \begin{equation}\label{eq:Mnk_norm}
        \|M_{n,k}\|_p = (k!)^{1/p}.
    \end{equation}
\end{Lemma}

\begin{IEEEproof}
Let~$e^1,\dots,e^{n}$ denote the canonical basis   of~$\R^{n}$.
Fix~$1\leq \ell_1<\dots<\ell_k\leq n$. Then~\eqref{eq:defmn}
yields 
\begin{align*} 
    L_{n,k} (e^{\ell_1} \otimes \dots \otimes e^{\ell_k}) &=  \begin{bmatrix} e^{\ell_1} & \cdots & e^{\ell_k} \end{bmatrix}^{(k)}. 
\end{align*}
This implies that~$|L_{n,k} x|_p \le |x|_p$ for any~$x\in\R^{ n^{k}}$, so 
    \begin{equation*}
        \|L_{n,k}\|_p  = \max_{x \in \R^{n^k} \setminus \{0\}} \frac{|L_{n,k}x|_p}{|x|_p} \le 1.
    \end{equation*}
    Furthermore,   
   $
        |e^1 \otimes \dots \otimes e^k|_p = 1,
  $  
    and
    \begin{equation*}
        |L_{n,k} (e^1 \otimes \dots \otimes e^k)|_p = |w^1|_p = 1,
    \end{equation*}
    where~$w^1$ is the first canonical vector of~$\R^{\binom{n}{k}}$. Therefore,~$\|L_{n,k}\|_p = 1$.

      To prove~\eqref{eq:Mnk_norm}, note that by~\eqref{eq:mnk_general}, 
\begin{align}
  &  M_{n,k} \begin{bmatrix}e^{\ell_1}  & \dots & e^{\ell_k}\end{bmatrix}^{(k)}\nonumber\\& = \sum_{(j_1,\dots,j_k) \in S(\ell_1,\dots,\ell_k)} \sigma(j_1,\dots,j_k) (e^{j_1} \otimes \dots \otimes e^{j_k}) .
\end{align}
This implies that for any~$y \in \R^{\binom{n}{k}}$ we have that~$M_{n,k}y$ is a vector that includes
every entry of~$y$ in~$k!$ entries, each  with either a plus or a minus sign. This implies~\eqref{eq:Mnk_norm}.
\end{IEEEproof}

\begin{figure*}[!t] 
\normalsize
        \begin{align*}
        \mu_{h,P_{n,m}}(\tilde J^{[2]}) &= \mu_h(P_{n,m} \tilde J^{[2]} P_{n,m}^T) \\
            &\le \mu \mleft( \begin{bmatrix}
            \mu_{p}(\tilde A^{[2]}) & \| M^T_{n,2}(I_n \otimes \tilde B)\|_{p} & 0 \\
            \|(I_n \otimes \tilde C)M_{n,2}\|_{p} & \mu_{p}(\tilde A \oplus \tilde D) & \|(\tilde B \otimes I_m)M_{m,2}\|_{p} \\
            0 & 
            \|M^T_{m,2}(\tilde C \otimes I_m)\|_{p} & \mu_{p}(\tilde D^{[2]})
        \end{bmatrix}  \mright).
    \end{align*}
\caption{Bounding the scaled hierarchic norm of~$\tilde J^{[2]}$.\label{fig:bound}}
\hrulefill
\vspace*{4pt}
\end{figure*}

\subsection{Proof of Theorem~\ref{thm:2cont_feedback}}\label{subsec:proofofmain}

We can now prove Theorem~\ref{thm:2cont_feedback}.
  Consider the feedback system~\eqref{eq:feedback}.
  Define  the state-variables~$\tilde x (t):= T_A x(t)$ and~$\tilde z (t):= T_D z(t)$. Then
    \begin{align*}
        \dot {\tilde x} &= T_A f(T_A^{-1} \tilde x, T_D^{-1} \tilde z), \\
        \dot {\tilde z} &= T_D g(T_A^{-1} \tilde x, T_D^{-1} \tilde z).
    \end{align*}
      The Jacobian of this system is the block matrix
    \begin{equation}
        \tilde J = \begin{bmatrix}
            \tilde A & \tilde B \\
            \tilde C & \tilde D
        \end{bmatrix},
    \end{equation}
with
\begin{align*}
        \tilde A(\tilde x,\tilde z) &:= T_A \frac{\partial f}{\partial x}(T_A^{-1} \tilde x, T_D^{-1} \tilde z) T_A^{-1}, \\
         \tilde B(\tilde x, \tilde z) &:= T_A \frac{\partial f}{\partial z}(T_A^{-1} \tilde x, T_D^{-1} \tilde z) T_D^{-1}, \\
        \tilde C(\tilde x, \tilde z) &:= T_D \frac{\partial g}{\partial x}(T_A^{-1} \tilde x, T_D^{-1} \tilde z) T_A^{-1}, \\
        \tilde D(\tilde x,\tilde z) &:= T_D \frac{\partial g}{\partial z}(T_A^{-1} \tilde x, T_D^{-1} \tilde z) T_D^{-1}.
    \end{align*}
 To bound a matrix measure of~$\tilde J^{[2]}$,     
  construct a hierarchic norm~$|\cdot|_h : \R^{\binom{n+m}{2}} \to \R_{\ge 0}$ as follows. Partition~$y\in \R^{\binom{n+m}{2}}$ as 
    \begin{equation*}
        y = \begin{bmatrix}
            y^1 \\
            y^2 \\
            y^3
        \end{bmatrix}, \quad \text{ with } y^1 \in \R^{\binom{n}{2}}, \ y^2 \in \R^{nm}, \ y^3 \in \R^{\binom{m}{2}}.
    \end{equation*}
    Fix a monotonic
vector norm~$|\cdot|_0:\R^3\to\R_{\geq 0}$, 
    and define the hierarchic norm 
    \begin{equation*}
        |y|_h : = \left| \begin{bmatrix}
            |y^1|_{p} \\
            |y^2|_{p} \\
            |y^3|_{p}
        \end{bmatrix} \right|_0.
    \end{equation*}
    Define also the scaled vector norm $|y|_{h,P_{n,m}} := |P_{n,m}y|_h$.
    Combining the expression in Fig.~\ref{fig:2add} and   
   Eq.~\eqref{eq:hier_bound} yields the bound in Fig.~\ref{fig:bound}.
   We   simplify 
   some of the terms in this bound as follows. First,
   \begin{align*}
       \mu_p\mleft(\tilde A^{[2]}\mright) & = \mu_p\mleft( (T_A \frac{\partial f}{\partial x} T_A^{-1})^{[2]} \mright)\\
      &= \mu_{p} \mleft(  T_A^{(2)} \frac{\partial f}{\partial x}^{[2]}  (T_A^{(2)})^{-1} \mright)  \\
       &= \mu_{p,T_A^{(2)}} \mleft(\frac{\partial f}{\partial x}^{[2]}\mright).
   \end{align*}
   Second, by Lemma~\ref{lem:AotimesB_pnorm}, 
    \begin{equation*}
        \mu_{p}(\tilde A \oplus \tilde D) = \mu_{p,T_A}\mleft(\frac{\partial f}{\partial x}\mright) + \mu_{p,T_D}\mleft(\frac{\partial g}{\partial z}\mright).
    \end{equation*}
Third, 
by Lemmas~\ref{lem:LnkMnk_norms} and~\ref{lem:AotimesB_pnorm}, we have
\begin{align*}
  \|  M^T_{n,2}(I_n\otimes \tilde B)\|_p &\leq \|M^T_{n,2}\|_p \| I_n\otimes \tilde B \|_p\\
  &=\|M_{n,2}\|_q \|\tilde B\|_p\\
  &=2^{1/q} \|\tilde B\|_p.
\end{align*}
Similarly, $\|(I_n \otimes \tilde C)M_{n,2}\|_p \le 2^{1/p} \|\tilde C\|_p$, $\|(\tilde B \otimes I_m) M_{m,2}\|_p \le 2^{1/p} \|\tilde B\|_p$, 
and~$|M^T_{m,2} (\tilde C\otimes I_m)|_p\leq 2^{1/q}\| \tilde C \|_p$. 
    Applying these expressions  to the bound in Fig.~\ref{fig:bound},
    we find that~$\mu_{h,P_{n,m}}(\tilde J^{[2]}) \leq \mu_0(S)$, 
    and this completes the proof of Theorem~\ref{thm:2cont_feedback}.~\hfill{\qed}

We believe that our results may find applications in various  feedback systems. The next section demonstrates this using a specific application. 

\section{An Application: $2$-contraction in a network of FitzHugh-Nagumo neurons}\label{sec:app}
  The FitzHugh-Nagumo neuron~(FHN) 
is a two-dimensional non-linear model designed to capture  important dynamic characteristics
of the   Hodgkin-Huxley model, including  oscillations, excitability, and
bistability (i.e., existence of 
two stable fixed points). It 
     has found applications in
neuroscience,  cardiac physiology, cell division, population dynamics,
and more; see the review paper~\cite{decades_fitzhugh_nagumo_model_guide}.

We apply the theoretical results to derive a sufficient condition for~$2$-contraction 
of  a dynamic  neural network that  includes~$N$ interconnected  FHNs.   
The~$i$th neuron, $i=1,\dots,N$, is modelled by the 2-dimensional system
\begin{equation}
    \begin{aligned}
        \dot v_i(t) &=c \left(  \omega_i (t)-\left( \frac{v_i^3(t)}{3}-v_i\right) +u_i \right) , \\
        \dot \omega_i(t) &= \frac{-1}{c } (v_i (t)- a  + b \omega_i(t)),\nonumber\\
        y_i(t)&=v_i(t),
    \end{aligned}
\end{equation}
where
\begin{enumerate}
    \item $v_i$ is the membrane voltage, diminished over time by a slower,   recovery variable~$w_i$;
    \item  $u_i$ [$y_i$] is the input [output] of neuron~$i$;   
    \item the parameters satisfy 
$a  ,b \geq 0 $, and~$c >0$.
\end{enumerate}

Let~$x:=\begin{bmatrix}
    v_1&\dots&v_N
\end{bmatrix}^T$,
and~$z:=\begin{bmatrix}
    \omega_1&\dots&\omega_N
\end{bmatrix}^T$. We assume that the input to the~$i$th neuron is a  weighted sum of all the membrane potentials. i.e., $u_i=(r^i)^T v$, with~$r^i \in \R^N $. Let~$R:=\begin{bmatrix}
   (r^1)^T\\\vdots\\(r^n)^T
\end{bmatrix}$. 
Then the neural network dynamics is
\begin{align}\label{eq:nagumo_total}
\dot x & =  c(z+x-\frac{1}{3}x^3 +R x),\nonumber\\
\dot z & = \frac{-1}{c} (  x-a 1_N +b z),
\end{align}
where we define
\[
x^3:=\begin{bmatrix}
    x_1^3&\dots&x_N^3,
\end{bmatrix},
\]
and~$1_N$ is the vector of~$N$ ones. 
The state-space is~$\Omega_x\times \Omega_z= \R^N \times \R^N$. 

Note that~\eqref{eq:nagumo_total}  has the form
of the feedback system~\eqref{eq:feedback} with
\begin{align*}
    f(x,z) &= c(z+x-\frac{1}{3}x^3 +R x),\\
    g(z,x) &= \frac{-1}{c} (  x-a 1_N +b z).  
 \end{align*}
Applying  
Corollary~\ref{coro:smallgain_2cont} allows to derive  
  a simple sufficient condition for 2-contraction  
of~\eqref{eq:nagumo_total}.

\begin{Corollary}\label{coro:net}
Fix~$p\in[1,\infty]$. If the following four conditions hold 
\begin{gather*}
    \mu_p(R^{[2]})<-2, \quad
      \mu_p(R)<  b c^{-2}-1,\quad 
      b>0,
    \\
      1<\frac{  (c^2(1+\mu_p(R)) -b) (2+\mu_p(R^{[2]} ))b }{2b-c^2(2+\mu_p(R^{[2]}))   } ,  
\end{gather*}
then the   FHN network~\eqref{eq:nagumo_total}
is~$2$-contractive. 
\end{Corollary}

 \begin{IEEEproof}
Let
\[
D(x):=\diag(x_1^2,\dots,x_N^2). 
\]
Then $
\frac{\partial f}{\partial x} = c ( I_N+R -D )
$,  $\frac{\partial f}{\partial z} = c I_N$, $
\frac{\partial g}{\partial x}=\frac{-1}{c} I_N $, and~$
\frac{\partial g}{\partial z} =  \frac{-b}{c} I_N $.  Hence,~$\frac{\partial f}{\partial x}^{[2]} = c ( 2I_r+R^{[2]} -D^{[2]} )   $, and~$ \frac{\partial g}{\partial z}^{[2]}
=  \frac{-2b}{c} I_r $,  with~$r:=\binom{N}{2} $.
Now calculating the quantities in Corollary~\ref{coro:smallgain_2cont}  
with $T_A = T_D = I_N$, and using well-known properties of matrix measures (see, e.g.~\cite{Desoer1972MatrixMeasure}), gives
\begin{align*}
 \Lip_z(f) &= c, \\
    \Lip_x(g) &= 1/c,\\
    \osL_x(f) &= c \sup_{x,z} \mu_p (I_N + R - D )  = c(1 + \mu_p(R)), \\
    \osL_z(g) &= -b/c, \\
\osL^{[2]}_x(f) &=c \sup_{x,z}   \mu_p (  
2I_r+R^{[2]}-D^{[2]}
 ) 
= c(2 + \mu_p(R^{[2]})), \\
\osL^{[2]}_z(g) &= -2b/c , 
\end{align*}
and the conditions in  Corollary~\ref{coro:smallgain_2cont}  
become the conditions in Corollary~\ref{coro:net}. 
 \end{IEEEproof}

The sufficient condition in Corollary~\ref{coro:net}   assumes a general connection matrix~$R$. If the network is diffusively coupled,  
that is,~$R=-L$ where $L$ is the Laplacian of an undirected graph, and we take~$p=2$ 
then the condition may be simplified.  Let~$\lambda_N\geq \dots \geq \lambda_2>\lambda_1=0$ denote the eigenvalues of~$L$, that is,~$\lambda_2$ is the algebraic connectivity of the associated  graph. Then~$\mu_2(R)=\mu_2(-L)=0$ and~$\mu_2(R^{[2]}) =
\mu_2(-L^{[2]})=-\lambda_2$. 
Then the 
condition for 2-contraction 
 in Corollary~\ref{coro:net},   becomes
\begin{gather*}
    \lambda_2 > 2, \quad 0 < bc^{-2} -1, \quad b>0, \\
    1 < \frac{(c^2 - b)(2 - \lambda_2)b}{2b-c^2(2 - \lambda_2)}.
\end{gather*}
The second and third conditions may be rewritten as 
\[
b > c^2>0,
\]
and the last condition is equivalent to: 
\begin{align*}
\frac{2}{2-\lambda_2} > c^2 b^{-1}+c^2-b. 
\end{align*}

\begin{figure}
    \centering
    \includegraphics[scale=0.65]{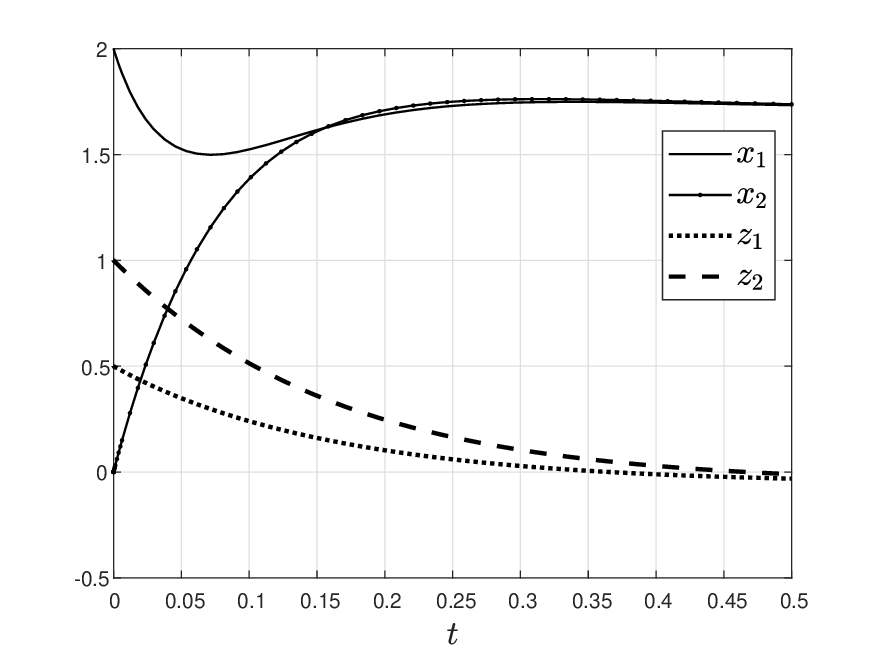}
    \caption{A trajectory of the FitzHugh-Nagumo network    in Example~\ref{exa:fhn_2_neurons}.}
    \label{fig:fhn_traj}
\end{figure}

\begin{example}\label{exa:fhn_2_neurons}
    Consider the case of a network with~$N=2$ FHNs, and parameters~$a=0$, $b=32$, $c=5$ and~$R=2\begin{bmatrix}
        -1&1\\1&-1
    \end{bmatrix}$.  
    Then~$R^{[2]}=\tr(R)=-4$, and  it is straightforward to verify that all the conditions in Corollary~\ref{coro:net} hold, so we conclude that~\eqref{eq:nagumo_total}
 is~$2$-contracting. 
Thus, any bounded solution converges to an equilibrium point. Fig.~\ref{fig:fhn_traj} 
depicts the solution of the network for the (arbitrarily chosen) initial
condition
$\begin{bmatrix}
   2&0&1/2&1
\end{bmatrix}^T$. 
Since~$a=0$, the origin is an equilibrium of the network, and it may be seen that the solution  converges to an equilibrium that is not the  origin. Hence, the network admits at least two equilibrium points, so in particular it is not contracting.
\end{example}
 
\section{Conclusion}
A contractive system admits a unique  exponentially stable equilibrium   and thus cannot model a multistationary system.
$k$-contraction  is a  generalization of contraction theory.  
In time-invariant $2$-contractive systems every bounded solution converges to the set of equilibria. This provides  a promising approach to the analysis and design of multistationary and multistable systems. Proving $k$-contraction is based on analyzing  the $k$-additive compound of the Jacobian of the vector field, and this is not trivial. 

Here, we derived a sufficient  condition for~$2$-contraction of a feedback system. This is based on deriving a new  expression for the~$2$-compounds of a block matrix, and combining this with a  hierarchic approach for proving $1$-contraction (i.e., standard contraction). We demonstrated the theoretical results by deriving a  simple
sufficient condition for~$2$-contraction in a dynamical neural network. 

An interesting topic for further research is control synthesis, that is, 
given a plant 
  design  a controller that guarantees $2$-contraction of the  closed-loop system. 

\bibliographystyle{IEEEtranS}
\bibliography{refs}

\end{document}